\documentclass[aps,prd,twocolumn,groupedaddress]{revtex4}
\usepackage{psfig}
\def\Journal#1#2#3#4{{#1} {\bf#2}, #3 (#4)}

\def\NPA{{\rm Nucl. Phys.} A}
\def\NPB{{\rm Nucl. Phys.} B}
\def\PLB{{\rm Phys. Lett.}  B}
\def\PRL{\rm Phys. Rev. Lett.}
\def\PRD{{\rm Phys. Rev.} D}
\def\PRC{{\rm Phys. Rev.} C}

\def\ep{\epsilon}
\def\vep{\varepsilon}
\def\la{\langle}
\def\ra{\rangle}

\def\al{\alpha}

\def\be{\begin{equation}}
\def\ee{\end{equation}}
\def\bea{\begin{eqnarray}}
\def\eea{\end{eqnarray}}
\input{epsfig.sty}
\begin{document}
\vspace{2.in}
\title{Transition Form Factors between Pseudoscalar and Vector Mesons
in Light-Front Dynamics}
\author{ Bernard L. G. Bakker$^{a}$, Ho-Meoyng Choi$^{b,c}$
and Chueng-Ryong Ji$^{d}$\\
$^a$ Department of Physics and Astrophysics, Vrije Universiteit,
     De Boelelaan 1081, NL-1081 HV Amsterdam, \\
     The Netherlands\\
$^b$ Department of Physics, Carnegie-Mellon University,
     Pittsburgh, PA 15213\\
$^c$ Department of Physics, Kyungpook National University,
Taegu, 702-701 Korea\\
$^d$ Department of Physics, North Carolina State University,
Raleigh, NC 27695-8202}

\begin{abstract}
We study the transition form factors between pseudoscalar and vector
mesons using a covariant fermion field theory model in $(3+1)$
dimensions. Performing the light-front calculation in the $q^+ =0$ frame in
parallel with the manifestly covariant calculation, we note that the
suspected nonvanishing zero-mode contribution to the light-front
current $J^+$ does not exist in our analysis of transition form
factors.  We also perform the light-front calculation in a purely
longitudinal $q^+ > 0$ frame and confirm that the form factors obtained
directly from the timelike region are identical to the ones obtained by
the analytic continuation from the spacelike region.  Our results for
the $B \rightarrow D^* l \nu_l$ decay process satisfy the constraints
on the heavy-to-heavy semileptonic decays imposed by the flavor
independence in the heavy quark limit.

\end{abstract}

\maketitle

\date{}

\section{Introduction}
{\label{sect.I}}
In a recent analysis of spin-one form factors in light-front dynamics,
we~\cite{BCJ} have shown that the zero-mode~\cite{ZM} complication can
exist even in the matrix element of the plus current $J^+$.  Using a
simple but exactly solvable model of the spin-one system with the
polarization vectors obtained from the light-front gauge
($\epsilon^+_{h=\pm 1} = 0$), we found that the zero-mode contribution
does not vanish in the helicity zero-to-zero amplitude. Neglecting the
zero-mode contribution results in the violation of angular
conditions~\cite{BJ}. There have been several recipes~\cite{GK,BH,CCKP}
in spin-one systems to extract the invariant form factors from the
matrix elements of the currents. Without taking into account the
zero-mode contribution, however, these different receipes do not
generate identical results in the physical form factors even if $J^+$
is used.

This indicates that the off-diagonal elements in the Fock-state
expansion of the current matrix cannot be neglected for the helicity
zero-to-zero amplitude even in reference frames where the plus
component of the momentum transfer, $q^+$, vanishes. Since the
factorization theorem in perturbative QCD (PQCD) relies essentially on
the helicity zero-to-zero matrix element diagonal in the Fock-state
expansion, the zero-mode contribution would complicate in principle the
PQCD analysis of the spin-one (and higher spin) systems. Fortunately,
our numerical computation indicates that the zero-mode contribution
diminishes significantly in the high momentum transfer region where the
PQCD analysis is applicable. Although the quantitative results that we
found from our model calculation may differ in other models depending
on the details of the dynamics in each model, the basic structure of
our calculation is common to any other model calculations including the
more phenomenological and realistic ones. Thus, we may expect the
essential findings from our model calculation to be supported further
by others. However, it doesn't preclude the possibility that the
zero-mode contribution may behave differently in different processes.
Thus, it appears important to analyze a different process involving a
spin-one system within the same model.

In this work, we analyze the transition form factors between
pseudoscalar and vector
mesons~\cite{Jaus90,Jaus96,MS,KT,OXT,CCH,Jaus,Jaus02}.  These form
factors can be measured in the semileptonic meson decay processes such
as $B \to D^* l \nu_l$ and $B \to \rho l \nu_l$ produced from
$B$-factories~\cite{JC}.  The physical region of momentum transfer
squared, $q^2$, for these processes (or form factors) is given by $4
m_l^2 \le q^2 \le (M_1 - M_2)^2$, where $M_1$ and $M_2$ are the masses
of the initial and final state mesons, respectively. This belongs to
the timelike region, while the elastic spin-one meson form factors
({\it i.e.}, $G_E, G_M, G_Q$) of for example the deuteron in the
electron deuteron elastic scattering experiment can only be measured in
the spacelike region, $q^2 \le 0$.  Not long ago, the same transition
form factors have been analyzed by Jaus~\cite{Jaus} using a light-like
four-vector called $\omega$ $(\omega^2=0)$ and the admixture of a
spurious $\omega$-dependent contribution was reported in the
axial-vector form factor $A_1(q^2)$ in the conventional light-front
formulas. The removal of the $\omega$-dependence in the physical form
factor amounts to the inclusion of the zero-mode contribution that we
present in this work.  However, the covariant formulation presented in
our work should be intrinsically distinguished from the formulation
involving $\omega$, since our formulation involves neither $\omega$ nor
any unphysical form factor.

This paper is organized as follows.  In Section \ref{sect.II}, we
present the manifestly covariant calculation of the transition form
factors between pseudoscalar and vector mesons using an exactly
solvable Bethe-Salpeter(BS) model of ($3+1$)-dimensional fermion field
theory.  In Section \ref{sect.III}, we apply the light-front dynamics
to calculate the same physical form factors. We separate the full
amplitudes into the valence and nonvalence contributions and compare
the results in the $q^+=0$ frame and the purely longitudinal $q^+ > 0$
frame. In the $q^+=0$ frame, we check whether the suspected zero-mode
contribution exists or not within our analysis.  In Section
\ref{sect.IV}, we present the numerical results for the transition form
factors making taxonomical decompositions of the full results into
valence and nonvalence contributions. Conclusions follow in Section
\ref{sect.V}.  In Appendix A, we summarize the kinematics of the
typical reference frames such as Drell-Yan-West (DWY), Breit (BRT), and
target-rest frame (TRF) in the transition form factor analysis. In
Appendix B, we present the manifestly covariant results of the
electromagnetic form factors and decay constants of the pseudoscalar
and vector mesons that are made of two unequal-mass constituents. These
results are used in fixing the model parameters of our numerical
analysis. In Appendices C and D, we present the more detailed formulae
used in the discussion of subsections \ref{SecIIID} and \ref{SecIIIE},
respectively.

\section{Manifestly Covariant Computation}
{\label{sect.II}}

The Lorentz-invariant transition form factors $g$, $f$, $a_{+}$, and
$a_{-}$ between a pseudoscalar meson with four-momentum $P_1$
and a vector meson with four-momentum $P_2$ and helicity $h$ are
defined~\cite{AW} by the matrix elements of the electroweak current
$J^\mu_{V-A} = V^\mu - A^\mu$ from the
initial state $|P_1;00\ra$ to the final state $|P_2;1h\ra$:
\begin{eqnarray}
&&\la P_2;1h|J^\mu_{V-A}|P_1;00\ra
\nonumber\\
&&\hspace{0.5cm} = 
i g(q^2) \varepsilon^{\mu\nu\alpha\beta} 
\epsilon^*_{\nu}P_\alpha q_\beta -f(q^2) \epsilon^{*\mu}
\nonumber\\
&&\hspace{0.5cm} -a_{+}(q^2)(\epsilon^{*}\cdot P)
P^\mu -a_{-}(q^2)(\epsilon^{*}\cdot P)q^\mu,
{\label{eq:1}}
\end{eqnarray}
where the momentum transfer $q^\mu$ is given by 
$q^\mu = P_1^\mu - P_2^\mu$, $P=P_1 + P_2$, 
and the polarization vector $\epsilon^*=\epsilon^*(P_2,h)$ of the 
final state vector
meson satisfies the Lorentz condition $\epsilon^* (P_2,h) \cdot P_2 = 0$.
While the form factor $g(q^2)$ is associated with the vector
current $V^\mu$, the rest of the form factors $f(q^2)$, $a_{+}(q^2)$, and
$a_{-}(q^2)$ are coming from the axial-vector current $A^\mu$. 
Thus, these transition form factors defined in Eq.~(\ref{eq:1}) are often 
given by the following convention~\cite{BSW},
\begin{eqnarray}
V(q^2) &=& (M_1 + M_2)g(q^2), \nonumber \\
A_1(q^2) &=& {f(q^2) \over {M_1 + M_2}}, \nonumber \\
A_2(q^2) &=& -(M_1 + M_2)a_{+}(q^2), \nonumber \\
A_0(q^2) &=& \frac{1}{2 M_2} \biggl[f(q^2)+ (M_1^2-M_2^2)a_{+}(q^2)
\nonumber \\
&+& q^2 a_{-}(q^2) \biggr],
{\label{eq:2}}
\end{eqnarray}
where $M_1$ and $M_2$ are the initial and final meson masses, respectively.

The solvable model, based on the covariant Bethe-Salpeter(BS)
model of ($3+1$)-dimensional fermion field theory, enables us to derive
the transition form factors between pseudoscalar and vector mesons
explicitly. The matrix element $\la P_2;1h|J^\mu_{V-A}|P_1;00\ra$ in this
model is given by
\begin{eqnarray}
&&\la P_2;1h|J^\mu_{V-A}|P_1;00\ra 
\nonumber\\
&&=
ig_1 g_2 \Lambda^2_1\Lambda^2_2\int\frac{d^4k}{(2\pi)^4}
\frac{S^{\mu\nu} \epsilon^*_\nu (P_2,h)}
{D_{\Lambda_1} D_{m_1} D_m D_{m_2} D_{\Lambda_2}},
{\label{eq:3}}
\end{eqnarray}
where $g_1$ and $g_2$ are the normalization factors which can be fixed
by requiring both charge form factors of pseudoscalar and vector mesons
to be unity at zero momentum transfer, respectively.  To regularize the
covariant fermion triangle-loop in ($3+1$) dimensions, we replace the
point gauge-boson vertex $\gamma^\mu (1-\gamma_5)$ by a non-local
(smeared) gauge-boson vertex
$\frac{{\Lambda_1}^2}{D_{\Lambda_1}}\gamma^\mu (1-\gamma_5)
\frac{{\Lambda_2}^2}{D_{\Lambda_2}}$, where $D_{\Lambda_1}
=(P_1-k)^2-{\Lambda_1}^2+i\vep$ and $D_{\Lambda_2}
=(P_2-k)^2-{\Lambda_2}^2+i\vep$, and thus the factor
$({\Lambda_1}{\Lambda_2})^2$ appears in the normalization factor.
$\Lambda_1$ and $\Lambda_2$ play the role of momentum cut-offs similar
to the Pauli-Villars regularization~\cite{BCJ1}.  The rest of the
denominators in Eq.~(\ref{eq:3}), {\it i.e.}, $D_{m_1} D_m D_{m_2}$,
are coming from the intermediate fermion propagators in the triangle
loop diagram and are given by
\begin{eqnarray}
D_{m_1} &=& (P_1 -k)^2 -{m_1}^2 + i\vep, \nonumber \\
D_m &=& k^2 - m^2 + i\vep, \nonumber \\
D_{m_2} &=& (P_2 -k)^2 -{m_2}^2 + i\vep.
{\label{eq:4}}
\end{eqnarray}
Furthermore, the trace term in Eq.~(\ref{eq:3}), $S^{\mu\nu}$, is given 
by
\begin{equation}
S^{\mu\nu} = {\rm Tr}[(\not\!p_2 + m_2)\gamma^\mu (1-\gamma_5)
(\not\!p_1 +m_1)\gamma_5(-\not\!k + m)\Gamma^\nu],
{\label{eqq:4}}
\end{equation}
where $m_1$, $m$, and  $m_2$ are the masses of the constituents
carrying the intermediate four-momenta $p_1=P_1 -k$, $k$, and $p_2=P_2
-k$, respectively.  For the vector meson vertex, we shall use
$\Gamma^\mu=\gamma^\mu$ in this section. While some modification of
this simple vertex will be considered in Section~\ref{SecIIIE}, our essential
findings are not altered by that modification.

Using the familiar trace theorems, we find for $S^{\mu\nu}$:
\begin{eqnarray}
S^{\mu\nu} = &&\hspace{-0.4cm} 4 i \vep^{\mu\nu\alpha\beta} [ k_\alpha
{P_1}_\beta (m-m_2) + k_\alpha {P_2}_\beta (m_1 -m) 
\nonumber\\
&+& {P_1}_\alpha {P_2}_\beta m ] \nonumber \\
&+& 4g^{\mu\nu} [m_1 k\cdot (k-P_2) + m_2 k\cdot(k- P_1) 
\nonumber\\
&-&m(k-P_1)\cdot (k-P_2) - {m_1}{m_2}m ] 
\nonumber \\
&+& 
4[ 2 k^\mu k^\nu (m-m_1) + k^\mu {P_1}^\nu (m_2 -m) 
\nonumber\\
&+& k^\mu {P_2}^\nu
(m_1 -m) - k^\nu {P_1}^\mu (m_2 +m)  
\nonumber \\
&+& k^\nu {P_2}^\mu (m_1 -m) + ({P_1}^\mu
{P_2}^\nu +{P_1}^\nu {P_2}^\mu)m ],
\nonumber\\
{\label{eq:5}}
\end{eqnarray}
where one should note that the ${P_2}^\nu$ terms will drop out once the
polarization vector ${\ep}^*_\nu (P_2,h)$ is multiplied into
$S^{\mu\nu}$. We have checked our result with the one obtained by
Jaus (see Eq.~(4.10) of Ref.~\cite{Jaus}) and found full agreement
between the two results.

We then decompose the product of five denominators given in Eq.~(\ref{eq:3}) 
into a sum of terms with three denominators only: {\it i.e.},
\bea
\frac{1}{D_{\Lambda_1} D_{m_1} D_m D_{m_2} D_{\Lambda_2}}
&=&
\frac{1}{({\Lambda_1}^2-{m_1}^2)({\Lambda_2}^2-{m_2}^2)}
\frac{1}{D_m}
\nonumber\\
&\times&\hspace{-0.2cm}
\biggl( \frac{1}{D_{\Lambda_1}} - \frac{1}{D_{m_1}}\biggr)
\biggl(\frac{1}{D_{\Lambda_2}} - \frac{1}{D_{m_2}} \biggr).
{\label{eq:6}}
\nonumber\\
\eea
Our treatment of the non-local smeared gauge-boson vertex
remedies~\cite{BCJ1} the conceptual difficulty associated with the
asymmetry appearing if the fermion-loop were regulated by smearing the
$q\bar{q}$ bound-state vertex. As discussed in our previous
work~\cite{BCJ1,BCJ}, the two methods lead to different results for the
calculation of the decay constants although they give the same result
for the form factors. For example, our result~\cite{BCJ} doesn't yield
a zero-mode contribution to the vector meson decay constant while the
asymmetric smearing of the hadronic vertex leads to the contamination
from the zero-mode~\cite{Jaus}.

Once we reduce the five propagators into a sum of terms containing
three propagators using Eq.~(\ref{eq:6}), we use the Feynman
parametrization for the three propagators, e.g.,
\begin{eqnarray}{\label{Feyn}}
&&\hspace{-0.5cm}\frac{1}{D_{m_1} D_m D_{m_2}}
\nonumber\\
&=&\hspace{-0.2cm}
\int^1_0 \hspace{-0.2cm}dx\int^{1-x}_0\hspace{-0.2cm} dy
\frac{2}{[D_m + (D_{m_1} - D_m) x + (D_{m_2}-D_m) y]^3}.
\nonumber\\
\end{eqnarray}
We then make a Wick rotation of Eq.~(\ref{eq:3}) in $D$-dimensions to
regularize the integral, since otherwise one looses the logarithmically
divergent terms in Eq.~(\ref{eq:3}).  Following the above procedure, we
finally obtain the Lorentz-invariant transition form factors as
follows:
\begin{widetext}
\begin{eqnarray}
g(q^2) &=& -\frac{{\cal N}}{8\pi^2}\int^1_0 dx\int^{1-x}_0 dy
[m_1 x + m_2 y +m(1-x-y)]C, \nonumber \\
f(q^2) &=& \frac{{\cal N}}{8\pi^2}\int^1_0 dx\int^{1-x}_0 dy
\biggl\{ 2(m_1-m+2m_2)
 \ln \biggl(\frac{C_{\Lambda_1 m_2}C_{m_1 \Lambda_2}}
{C_{\Lambda_1 \Lambda_2}C_{m_1 m_2}} \biggr)
\nonumber \\
&+& \biggl[ 2(m_1 + m_2 -m) \{ (x+y)(xM_1^2 +yM_2^2)-xyq^2\}
-m_1\{2yM_2^2 + x(M_1^2 + M_2^2 -q^2)\}
\nonumber \\
&-&m_2\{2xM_1^2 + y(M_1^2 + M_2^2 -q^2)\}
+ m\{2xM_1^2 + 2yM_2^2 +(x+y-1)(M_1^2 + M_2^2 -q^2)\} - 2m_1 m_2 m
\biggr]C
\biggr\}, 
\nonumber \\
a_{+}(q^2) &=& \frac{{\cal N}}{8\pi^2}\int^1_0 dx\int^{1-x}_0 dy
[ (x+y)\{2x(m-m_1)+m_2-m\} +x(m_1-m_2-2m)+m ]C, \nonumber \\
a_{-}(q^2) &=& \frac{{\cal N}}{8\pi^2}\int^1_0 dx\int^{1-x}_0 dy
[
(x-y)\{2x(m-m_1)+m_2-m\}-x(m_1+m_2)-m
]C,
{\label{eq:7}}
\end{eqnarray}
\end{widetext}
where
${\cal N}= g_1 g_2\Lambda^2_1\Lambda^2_2/ 
({\Lambda_1}^2-{m_1}^2)({\Lambda_2}^2-{m_2}^2)$ and
$C=(1/C_{\Lambda_1 \Lambda_2} -1/C_{\Lambda_1 m_2} - 1/C_{m_1 \Lambda_2} +
1/C_{m_1 m_2})$ with
\begin{eqnarray}
C_{\Lambda_1 \Lambda_2}&=& (1-x-y)(x M_1^2 + y M_2^2) + xy q^2 
\nonumber\\
&&- (x \Lambda_1^2 + y \Lambda_2^2) - (1-x-y)m^2, \nonumber \\
C_{\Lambda_1 m_2}&=& (1-x-y)(x M_1^2 + y M_2^2) + xy q^2 
\nonumber\\
&&- (x \Lambda_1^2 + y m_2^2) - (1-x-y)m^2, \nonumber \\
C_{m_1 \Lambda_2}&=& (1-x-y)(x M_1^2 + y M_2^2) + xy q^2 
\nonumber\\
&&- (x m_1^2 + y \Lambda_2^2) - (1-x-y)m^2, \nonumber \\
C_{m_1 m_2}&=& (1-x-y)(x M_1^2 + y M_2^2) + xy q^2 
\nonumber\\
&&- (x m_1^2 + y m_2^2) - (1-x-y)m^2. 
{\label{eq:8}}
\end{eqnarray}
Note that the logarithmic term in $f(q^2)$ is obtained from the
dimensional regularization with the Wick rotation.

\section{Light-front calculation}
{\label{sect.III}}
It is native to the light-front analysis that a judicious choice of the 
current component is important for an effective computation of matrix 
elements. For the present work, we shall use only the plus-component of 
the current matrix element $\la P_2;1h|J^\mu_{V-A}|P_1;00\ra$
in the calculation of the transition form factors.  

As we did in Ref.~\cite{BCJ}, the LF calculation for 
the trace term in Eq.~(\ref{eqq:4}) with plus current ($\mu=+$)
can be separated into the on-shell propagating part $S^{+}_{\rm on}$
and the instantaneous part $S^{+}_{\rm inst}$ via
\bea{\label{sep}}
\not\!p + m &=&
(\not\!p_{\rm on} + m) + \frac{1}{2}\gamma^+(p^- - p^-_{\rm on})
\eea
as
\bea{\label{ch:0}}
S^+_h &=& S^{+\nu}  \epsilon^*_\nu(P_2,h)
= S^{+(h)}_{\rm on} + S^{+(h)}_{\rm inst},
\eea
where
\bea{\label{ch:1}}
S^{+(h)}_{\rm on\;V-A}\hspace{-0.1cm} &=& \hspace{-0.1cm}
-4i\varepsilon^{+\mu\nu\al}
[ m_1(p_{2\rm on})_\mu (k_{\rm on})_\nu
- m_2 (p_{1\rm on})_\mu (k_{\rm on})_\nu
\nonumber\\
&&\;\hspace{1.2cm}
- m (p_{1\rm on})_\mu (p_{2\rm on})_\nu ]\epsilon^*_\al
\nonumber\\
&& + 4 m_1[ (k_{\rm on}\cdot\epsilon^*)p^+_{2\rm on}
+ (p_{2\rm on}\cdot\epsilon^*)k^+_{\rm on}
\nonumber\\
&&\;\hspace{1.2cm}- (p_{2\rm on}\cdot k_{\rm on})\epsilon^{*+} ]
\nonumber\\
&& - 4 m_2[ (k_{\rm on}\cdot\epsilon^*)p^+_{1\rm on}
- (p_{1\rm on}\cdot\epsilon^*)k^+_{\rm on}
\nonumber\\
&&\;\hspace{1.2cm} + (p_{1\rm on}\cdot k_{\rm on})\epsilon^{*+} ]
\nonumber\\
&& + 4 m[ (p_{2\rm on}\cdot\epsilon^*)p^+_{1\rm on}
+ (p_{1\rm on}\cdot\epsilon^*)p^+_{2\rm on}
\nonumber\\
&&\;\hspace{1.2cm}- (p_{1\rm on}\cdot p_{2\rm on})\epsilon^{*+} ]
\nonumber\\
&&- 4 m_1 m_2 m \epsilon^{*+},
\eea
and
\bea{\label{ch:2}}
S^{+(h)}_{\rm inst\;V-A}
&=& -4(k^- -k^-_{\rm on})m_2 p^+_{1\rm on}\epsilon^{*+},
\eea
with $p_1=P_1 -k$, $p_2=P_2 -k$. The subscript (on) denotes the on-mass
shell ($p^2=m^2$) quark momentum, {\it i.e.}, $p^-=p^-_{\rm on} =
(m^2+{\bf p}^2_\perp)/p^+$.  Note that the first term of $S^+_{\rm on}$
corresponds to the vector current matrix element and the rest to the
axial-vector current matrix element. The instantaneous contribution
$S^+_{\rm inst}$ comes only from the axial-vector current, {\it i.e.},
$S^{+(h)}_{\rm inst\;V-A}=-S^{+(h)}_{\rm inst\;A}$.

The polarization vectors used in this analysis are given by
\bea{\label{pol_vec}}
\epsilon^\mu(\pm 1)&=&[\ep^+,\ep^-,\ep_\perp]
=\biggl[0,\frac{2}{P^+_2}{\bf\epsilon}_\perp(\pm)\cdot{\bf P_{2\perp}},
{\bf\epsilon}_\perp(\pm 1)\biggr],
\nonumber\\
{\bf\epsilon}_\perp(\pm 1)&=&\mp\frac{(1,\pm i)}{\sqrt{2}},
\nonumber\\
\epsilon^\mu(0)&=&
\frac{1}{M_2}\biggl[P^+_2,\frac{{\bf P}^2_{2\perp}-M^2_2}{P^+_2},
{\bf P}_{2\perp}\biggr].
\eea
The traces in Eqs. (\ref{ch:1}) and (\ref{ch:2}) are then obtained as 
\bea{\label{pol:1}}
S^{+(h=1)}_{\rm on\;V} 
&=&\hspace{-0.2cm}\frac{2P^+_1}{\sqrt{2}}\varepsilon^{+-xy}
\biggl\{ q^L {\cal A}_P
+k^L\;[\; (m - m_2)(1-x) 
\nonumber\\
&&+ (m_1 -m)(\al -x) + (m_1 -m_2)x \;]
\biggr\},
\nonumber\\
S^{+(h=1)}_{\rm on\;A}
&=&\hspace{-0.2cm}-\frac{4P^+_1}{\sqrt{2}}
\biggl\{ 
\frac{(\al - 2x)}{\al}q^L {\cal A}_P
\nonumber\\
&&+ k^L\; [\; (\al - 2x)(m_1 -m) - (m_2 + m)\;]
\biggr\},
\nonumber\\
S^{+(h=1)}_{\rm inst\;V}&=&S^{+(h=1)}_{\rm inst\;A}=0,
\eea
for the transverse polarization vector ($h=1$) and 
\bea{\label{pol:2}}
S^{+(h=0)}_{\rm on\;A}
&=&\hspace{-0.2cm}\frac{4P^+_1}{x' M_2}
\biggl\{
{\cal A}_P[\; x'(1-x')M^2_2 + m_2 m + x'^2{\bf q}^2_\perp\;]
\nonumber\\
&&+{\bf k}^2_\perp (x m_1 + m_2 - x m)
\nonumber\\
&&
+ x' {\bf k}_\perp\cdot{\bf q}_\perp
[\; 2x(m_1 -m) + m_2 + m\;]
\biggr\},
\nonumber\\
S^{+(h=0)}_{\rm inst\;A}&=&
\frac{4\al(P^+_1)^2}{M_2}(1-x)m_2(k^- - k^-_{\rm on}),
\nonumber\\
\eea
for the longitudinal one ($h=0$), where 
\begin{eqnarray}
 \al & = & P^+_2/P^+_1=1-q^+/P^+_1, \; x = k^+/P^+_1,
 \; x'=x/\al, \nonumber \\
 q^L & = & q_x -iq_y, \;  k^L = k_x -ik_y, \nonumber \\
 {\cal A}_P & =  & x m_1 + (1-x) m.
\label{eq.xx}
\end{eqnarray}
Here, we used the ${\bf P}_{1\perp}=0$ frame.  The
(timelike) momentum transfer $q^2=(P_1-P_2)^2$ is in general given by
\bea{\label{tq2}}
q^2&=&q^+q^- - {\bf q}^2_\perp =(1-\al)\biggl(M^2_1 - 
\frac{M^2_2}{\al}\biggr) -\frac{{\bf q}^2_\perp}{\alpha}.
\eea 

Defining the matrix element 
$\la P_2;1h|J^+_{V-A}|P_1;00\ra\equiv\la J^+_{V-A}\ra^h$
of the plus component of the V--A current in 
Eq.~(\ref{eq:3})  as
\begin{equation}{\label{ch:3}}
\la J^+_{V-A}\ra^h=
ig_1g_2\Lambda^2_1\Lambda^2_2\int\frac{d^4k}{(2\pi)^4}
\frac{(S^{+(h)}_{\rm on} + S^{+(h)}_{\rm inst})_{V-A}}
{D_{\Lambda_1} D_{m_1} D_m D_{m_2} D_{\Lambda_2}},
\end{equation}
one obtains the relations between the current matrix elements and
the weak form factors as follows 
\bea{\label{ch:4}}
\la J^+_V\ra^{h=1} &=& 
-\frac{P^+_1}{\sqrt{2}}\varepsilon^{+-xy}q^L g(q^2),
\nonumber\\
\la J^+_V\ra^{h=0} &=& 0,
\eea
for the vector current and
\bea{\label{ch:5}}
\la J^+_A\ra^{h=1} 
&=& \frac{P^+_1q^L}{\al\sqrt{2}}
\biggl[ (1+\al) a_+(q^2) 
+ (1-\al) a_-(q^2)\biggr],
\nonumber\\
\la J^+_A\ra^{h=0} &=& \frac{\al P^+_1}{M_2}f(q^2)
+ \frac{\al P^+_1}{2M_2}
\biggl(M^2_1 - \frac{M^2_2}{\al^2} 
+ \frac{{\bf q}^2_\perp}{\al^2}\biggr)
\nonumber\\
&\times&\biggl[ (1+\al)a_+(q^2) + (1-\al)a_-(q^2)\biggr],
\eea
for the axial-vector current. 

\subsection{Methods of extracting weak form factors}
\label{SecIIIA}
The extraction of weak form factors can be done in various ways. Among
them, there are two popular ways of extracting the form factors, {\it
i.e.}, (1) the form factors are obtained in the spacelike region using
the $q^+ =0$ frame and then analytically continued to the timelike
region by changing ${\bf q}_\perp$ to $i{\bf q}_\perp$, (2) the form
factors are obtained by a direct timelike analysis using a $q^+>0$
frame. In this work, we shall analyze the form factors in both ways.

In the $q^+=0$ frame ({\it i.e.}, $\al=1$) with the transverse
polarization modes, one could extract the form factors $g(q^2)$ and
$a_+(q^2)$ without including the zero-mode contributions as one can see
from Eqs.~(\ref{ch:4}) and~(\ref{ch:5}).  One could in principle obtain
the form factor $f(q^2)$ in the $q^+=0$ frame and the longitudinal
polarization mode.  In this case, it is important to check whether the
zero-mode contribution exists or not by investigating the instantaneous
part of the trace given by Eq.~(\ref{pol:2}).  In particular, as we
discussed in Section \ref{sect.I}, the admixture of spurious
$\omega$-dependent contributions was reported~\cite{Jaus} indicating a
possible zero-mode contribution to the axial form factor $A_1(q^2)$
which is essentially identical to $f(q^2)$ modulo some constant factor
(see Eq.~(\ref{eq:2})).  As we shall show in Subsections \ref{SecIIID}
and \ref{SecIIIE}, however, we find that the zero-mode contribution to
the form factor $f(q^2)$ does not exist in our analysis.

Using only the plus current $J_{V-A}^+$ in the $q^+ = 0$ frame,
it is not possible to extract the form factor $a_-(q^2)$. 
On the other hand, if one chooses a $q^+>0$ frame, specifically
a purely longitudinal momentum frame where the momentum transfer
is given by 
\bea{\label{pl:1}}
q^2&=&q^+q^- =(1-\al)\biggl(M^2_1 - \frac{M^2_2}{\al}\biggr),
\eea
one can extract all four form factors by using only the plus-current.
We compute them all in this purely longitudinal momentum frame
including the nonvalence contributions for the matrix elements.
This frame corresponds to the case $\theta =0$ or $\pi$ in the TRF and BRT 
frames summarized in Appendix A.

For this particular choice of the purely longitudinal frame,
there are two solutions of $\al$ for a given $q^2$, {\it i.e.},
\begin{equation}{\label{pl:2}}
\al_{\pm}=\frac{M_2}{M_1}\biggl[
\frac{M^2_1 + M^2_2-q^2}{2M_1M_2} \pm
\sqrt{ \biggl(\frac{M^2_1 + M^2_2-q^2}{2M_1M_2}\biggr)^2-1}
\biggr],
\end{equation}
where the $+(-)$ sign in Eq.~(\ref{pl:2}) corresponds to the daugther
meson recoiling in the positive(negative) $z$-direction relative to
the parent meson. At zero recoil ($q^2=q^2_{\rm max}$) and maximum
recoil ($q^2=0$), $\alpha_\pm$ are given by
\bea{\label{apm}}
\alpha_+(q^2_{\rm max})&=&\alpha_-(q^2_{\rm max})=\frac{M_2}{M_1},
\nonumber\\
\alpha_+(0)&=&1,\;\; 
\alpha_-(0)=\biggl(\frac{M_2}{M_1}\biggr)^2.
\eea
The form factors are of course independent of the recoil directions
($\alpha_\pm$) if the nonvalence contributions are added to the valence
ones.  As one can see from Eqs. (\ref{ch:4}) and (\ref{ch:5}), however,
one should be careful in setting ${\bf q}_\perp = 0$ to get the results
in this frame.  One cannot simply set ${\bf q}_\perp=0$ from the start,
but may set it to zero only after the form factors are extracted.

While the form factor $g(q^2)$ in the $q^+>0$ frame can be obtained
directly from Eq.~(\ref{ch:4}), the form factor $f(q^2)$ can be
obtained only after $a_\pm(q^2)$ are calculated.

To illustrate this, we define
\bea{\label{pl:3}}
\la J^+_A\ra^{h=1} |_{\al=\al_\pm}
&\equiv& \frac{P^+_1 q^L}{\sqrt{2}}I^+_A(\al_\pm),
\eea
and obtain from Eq.~(\ref{ch:5}) 
\bea{\label{pl:4}}
a_+(q^2)&=&\frac{ 
\al_+(1-\al_-)I^+_A(\al_+) - \al_-(1-\al_+) I^+_A(\al_-)}
{2(\al_+ -\al_-)},
\nonumber\\
a_-(q^2)&=&-\frac{
\al_+(1+\al_-)I^+_A(\al_+) - \al_-(1+\al_+) I^+_A(\al_-)}
{2(\al_+ -\al_-)},
\nonumber\\
\eea
and
\bea{\label{pl:5}}
f(q^2)&=& \frac{M_2}{\al P^+_1}\la J^+_A\ra^{h=0}
-\frac{1}{2}\biggl(M^2_1-\frac{M^2_2}{\al^2}\biggr)
\nonumber\\
&&\times\biggl[(1+\al)a_+(q^2) + (1-\al)a_-(q^2) \biggr].
\eea

\subsection{Valence contribution to $\la J^+_{V-A}\ra^h$}
\label{SecIIIB}
In the valence region $0<k^+<P^+_2$, the pole $k^-=k^-_{\rm on}=(m^2 +
{\bf k}^2_\perp -i\vep)/k^+$ ({\it i.e.}, the spectator quark) is located in
the lower half of the complex $k^-$-plane.  Thus, the Cauchy integration
formula for the $k^-$-integral in Eq.~(\ref{ch:3}) gives 
\begin{widetext}
\bea{\label{ch:6}}
\la J^{+}_{V-A}\ra^h_{\rm val}&=& 
\frac{g_1g_2\Lambda^2_1\Lambda^2_2}{2(2\pi)^3}\int^\al_0
\frac{dx}{x(1-x)^2(1-x')^2}\int d^2{\bf k}_\perp
\frac{S^{+(h)}_{\rm on\;V-A}}
{(M^2_1 - M^2_0)(M^2_1-M^2_{\Lambda_1})
(M^2_2-M'^2_0)(M^2_2-M'^2_{\Lambda_2})},
\eea 
\end{widetext}
where
\bea{\label{ch:7}}
M^2_0 &=& \frac{{\bf k}^2_\perp + m^2_1}{1-x}
+ \frac{{\bf k}^2_\perp + m^2}{x},
\nonumber\\
M'^2_0 &=& \frac{{\bf k'}^2_\perp + m^2_2}{1-x'}
+ \frac{{\bf k'}^2_\perp + m^2}{x'},
\eea
and 
$M^2_{\Lambda_1}=M^2_0(m_1\to\Lambda_1)$,
$M'^2_{\Lambda_2}= M'^2_0(m_2\to\Lambda_2)$ 
with ${\bf k'}_\perp={\bf k}_\perp + x'{\bf q}_\perp$. 
Note that there is no instantaneous contribution in the valence region.
From Eqs.~(\ref{pol:1}) and (\ref{ch:4}), we obtain the valence contribution
to $g(q^2)$ as follows
\begin{widetext}
\bea{\label{g_val}}
g(q^2)_{\rm val}&=& 
-\frac{g_1g_2\Lambda^2_1\Lambda^2_2}{\al(2\pi)^3}\int^\al_0
\frac{dx}{x(1-x)^2(1-x')^2}\int d^2{\bf k}_\perp
\frac{1}{(M^2_1 - M^2_0)(M^2_1-M^2_{\Lambda_1})
(M^2_2-M'^2_0)(M^2_2-M'^2_{\Lambda_2})}
\nonumber\\
&&\times\biggl\{ {\cal A}_P +
\frac{{\bf k}_\perp\cdot{\bf q}_\perp}{{\bf q}^2_\perp}
\;[\; (m - m_2)(1-x) + (m_1 -m)(\al -x) + (m_1 -m_2)x \;] \biggr\}.
\eea
\end{widetext}
While Eq.~(\ref{g_val}) accounts only for the valence contribution in
the $q^+>0$ frame, it is the exact solution in the $q^+=0$ ({\it i.e.},
$\al=1$) frame due to the absence of the zero-mode contribution. Here,
we should note the discrepancy between Ref.~\cite{Jaus90} and
Refs.~\cite{OXT,CCH} for the calculation of the $g(q^2)$ form factor.
For the simple vector meson vertex of $\Gamma^\mu = \gamma^\mu$, our
result is the same as Ref.~\cite{Jaus90} but different from
Refs.~\cite{OXT,CCH}.  The authors of  Refs.~\cite{OXT,CCH} claimed to
compute the ``$+$'' component of the vector current (see for instance
Eq.~(2.75) in ~\cite{CCH}).  However, they indeed used the ``$-$''
component of the current instead of the ``$+$'' one. In their
computation they used the coefficient of $\epsilon_{+-xy}$ which
corresponds to $\gamma^-$ for the electroweak current vertex rather
than the coefficient of $\epsilon_{-+xy}$(or equivalently
$\epsilon^{+-xy}$) that corresponds to the plus current.  This
difference in choosing the component of the current caused the
discrepancy between the results of Ref.~\cite{Jaus90} and
Refs.~\cite{OXT,CCH}.
It is well known \cite{BCJ1} that the minus current contains zero-mode
contributions.

In the $q^+=0$ frame, the valence contribution to $a_+(q^2)$ is the exact
solution, again due to the absence of the zero-mode contribution.  The
result is obtained from Eqs.~(\ref{pol:1}) and (\ref{ch:5}) as

\begin{widetext} 
\bea{\label{ap_val}} a_+(q^2)|_{q^+=0}&=&
-\frac{g_1g_2\Lambda^2_1\Lambda^2_2}{(2\pi)^3}\int^1_0
\frac{dx}{x(1-x)^4}\int d^2{\bf k}_\perp \frac{1}{(M^2_1 -
M^2_0)(M^2_1-M^2_{\Lambda_1}) (M^2_2-M'^2_0)(M^2_2-M'^2_{\Lambda_2})}
\nonumber\\ &&\times\biggl\{ (1-2x){\cal A}_P + \frac{{\bf
k}_\perp\cdot{\bf q}_\perp}{{\bf q}^2_\perp} \;[\; (1-2x) m_1 - m_2 -
2(1-x) m \;] \biggr\}.  \eea \end{widetext}

As we have shown in the present subsection, \ref{SecIIIB}, the two form factors
$g(q^2)$ and $a_+(q^2)$ can be computed in the $q^+ = 0$ frame. The form 
factor $f(q^2)$ can also be computed in the same frame, as we discussed
in the last subsection \ref{SecIIIA}. The lack of a zero-mode contribution to
$f(q^2)$ is discussed in the subsections \ref{SecIIID} and \ref{SecIIIE}.
Before we discuss this point, we first complete the presentation of 
the matrix element, {\it i.e.},
\bea{\label{totC}}
\la J^+_{V-A}\ra^h &=& 
\la J^+_{V-A}\ra^h_{\rm val} + \la J^+_{V-A}\ra^h_{\rm nv},
\eea
by computing the nonvalence contribution $\la J^+_{V-A}\ra^h_{\rm nv}$
in the next subsection, \ref{SecIIIC} for an arbitrary $q^+$(or $\al$) value.
The nonvalence contribution is necessary to compute the form factors in
the purely longitudinal $q^+ > 0$ frame. It is confirmed in our
numerical results (Section IV) that the values of the calculated form
factors in the $q^+=0$ frame are identical to those in the purely
longitudinal $q^+ > 0$ frame, as they should be when the nonvalence
contribution is added to the valence one. In the purely longitudinal
$q^+>0$ frame, we shall use Eqs.~(\ref{pl:4}) and (\ref{pl:5}) to
obtain the form factors $a_{\pm}(q^2)$ and $f(q^2)$, while the form
factor $g(q^2)$ can be obtained directly from Eq.~(\ref{ch:4}).

\subsection{Nonvalence contribution to $\la J^+_{V-A}\ra^h$}
\label{SecIIIC}
In the nonvalence region $P^+_2<k^+<P^+_1$, the poles are at
$k^- = k^-_{m_1} \equiv
P^-_1 + [m^2_1 +({\bf k}_\perp -{\bf P}_{1\perp})^2 -i\vep]/(k^+-P^+_1)$
(from the struck quark propagator) and
$k^- = k^-_{\Lambda_1} \equiv P^-_1 
+ [\Lambda^2_1+({\bf k}_\perp -{\bf P}_{1\perp})^2 -i\vep]/(k^+-P^+_1)$
(from the smeared quark-photon vertex),
and are located in the upper half of the complex $k^-$-plane.

When we do the Cauchy integration over $k^-$ to obtain the LF
time-ordered diagrams, we use Eq.~(\ref{eq:6}) to avoid the complexity 
of treating double $k^-$-poles and obtain
\begin{widetext}
\bea{\label{jnv}}
\la J^+_{V-A}\ra^h_{\rm nv}
&=&\frac{{\cal N}}{2(2\pi)^3}
\int^1_\al\frac{dx}{xx''(x-\al)}\int d^2{\bf k}_\perp
\biggl\{
\frac{S^{+(h)}_{\rm on} + S^{+(h)}_{\rm inst}(k^-=k^-_{\Lambda_1})}
{(M^2_1-M^2_{\Lambda_1})(q^2-M^2_{\Lambda_1\Lambda_2})}
- \frac{S^{+(h)}_{\rm on} + S^{+(h)}_{\rm inst}(k^-=k^-_{\Lambda_1})}
{(M^2_1-M^2_{\Lambda_1})(q^2-M^2_{\Lambda_1m_2})}
\nonumber\\
&&+ \frac{S^{+(h)}_{\rm on} + S^{+(h)}_{\rm inst}(k^-=k^-_{m_1})}
{(M^2_1-M^2_{0})(q^2-M^2_{m_1m_2})}
- \frac{S^{+(h)}_{\rm on} + S^{+(h)}_{\rm inst}(k^-=k^-_{m_1})}
{(M^2_1-M^2_{0})(q^2-M^2_{m_1\Lambda_2})}
\biggr\},
\eea
\end{widetext}
where $M^2_{\Lambda_1}$ is defined just below Eq.~(\ref{ch:7}) and
\bea{\label{jnv_M}}
M^2_{\Lambda_1\Lambda_2}&=&
\frac{{\bf k''}^2_\perp + \Lambda^2_1}{x''}
+ \frac{{\bf k''}^2_\perp + \Lambda^2_2}{1-x''}, 
\nonumber\\
M^2_{\Lambda_1m_2}&=&
\frac{{\bf k''}^2_\perp + \Lambda^2_1}{x''}
+ \frac{{\bf k''}^2_\perp + m^2_2}{1-x''},
\nonumber\\
M^2_{m_1m_2}&=&
\frac{{\bf k''}^2_\perp + m^2_1}{x''}
+ \frac{{\bf k''}^2_\perp + m^2_2}{1-x''}, 
\nonumber\\
M^2_{m_1\Lambda_2}&=&
\frac{{\bf k''}^2_\perp + m^2_1}{x''}
+ \frac{{\bf k''}^2_\perp + \Lambda^2_2}{1-x''},
\eea
with the variables defined by
\bea{\label{jnv_xk}}
x''&=& \frac{1-x}{1-\al},\;\;
{\bf k''}_\perp = {\bf k}_\perp + x''{\bf q}_\perp.
\eea
Note that the instantaneous contribution $S^{+(h)}_{\rm inst}(k^-)$ in
Eq.~(\ref{jnv}) exists only for the longitudinal polarization vector
case ($h=0$).  The total current matrix element is then given by
Eq.~(\ref{totC}).

\subsection{Is the form factor $f(q^2)$ immune to the
zero-mode in the $q^+=0$ frame?}
\label{SecIIID}
Using the plus component of the axial-current given by
Eq.~(\ref{ch:5}), the form factor $f(q^2)$ is obtained from the mixture
of the longitudinal polarization vector ({\it i.e.}, $\la
J^+_A\ra^{h=0}$) and the transverse one ({\it i.e.}, $\la
J^+_A\ra^{h=1}$).

Especially, in the $q^+=0$ frame ({\it i.e.}, the $\alpha\to 1$ limit),
the form factor $f(q^2)$ is given by
\bea{\label{f_val}}
 f(q^2)= -(M^2_1 - M^2_2 + {\bf q}^2_\perp)a_+(q^2)
 +\frac{M_2}{P^+_1}\la J^+_A\ra^{h=0},
\eea
where $a_+(q^2)$ is given by Eq~(\ref{ap_val}) and the
valuence contribution to $\la J^+_A\ra^{h=0}$ in the $q^+=0$ frame is
given by
\begin{widetext}
\bea\label{JAh0_val}
\la J^+_A\ra^{h=0}_{\rm val}&=&
\frac{2P^+_1 g_1 g_2 \Lambda_1\Lambda_2}{(2\pi)^3 M_2}
\int^1_0\frac{dx}{x(1-x)^4}\int d^2{\bf k}_{\perp}
\frac{1}{(M^2_1-M^2_0)(M^2_1-M^2_{\Lambda_1})
(M^2_2-M'^2_0)(M^2_2-M'^2_{\Lambda_2})}
\nonumber\\
&&\;\times
\biggl\{
{\cal A}_p[(1-x)M^2_2 + \frac{m_2m}{x}+ x {\bf q}^2_\perp]
 + {\bf k}^2_\perp(m_1 + \frac{m_2}{x} - m)
+ {\bf k}_\perp\cdot{\bf q}_\perp[2x(m_1-m)+m_2+m]
\biggr\}.
\eea
\end{widetext}
The zero-mode contribution is obtained from the
$\alpha\to 1$ limit of $\la J^+_A\ra^{h}_{nv}$ in Eq.~(\ref{jnv}).
As the only possible source for the zero-mode is the factor $k^- -
k^-_{on}$ appearing in Eq.~(\ref{ch:2}), only the instantaneous parts
of the trace terms could be the origin of a zero-mode contribution.
Since $S^{+(h=1)}_{\rm inst\; A}=0$, the form factor $a_+(q^2)$ 
is immune to the zero-mode. Thus, we only need to check the zero-mode 
contribution to the matrix element of $\la J^+_A\ra^{h=0}$ 
using Eq.~(\ref{jnv}).
 
The zero-mode contribution(if it exists) to $\la J^+_A\ra^{h=0}$ in 
Eq.~(\ref{jnv}) is proportional to
\bea{\label{fz:1}}
I^{\rm zm}_A&\sim&
\lim_{\alpha\to 1}\int^1_\alpha
\frac{dx d^2{\bf k}_\perp}{xx''(x-\alpha)}
\frac{S^{+(h=0)}_{\rm inst}(k^-=k^-_{\Lambda_1})}
{(M^2_1-M^2_{\Lambda_1})
(q^2-M^2_{\Lambda_1\Lambda_2})}
\nonumber\\
&&+\cdots,
\eea
where ($\cdots$) represent the other three instantaneous terms in 
Eq.~(\ref{jnv})
and $S^{+(h=0)}_{\rm inst}(k^-)$ is given by Eq.~(\ref{pol:2}).

Showing only the longitudinal momentum fraction factors relevant to the 
zero-mode, one can easily find that Eq.~(\ref{fz:1}) becomes 
\bea{\label{fz:2}}
I^{\rm zm}_A &\sim& \lim_{\alpha\to 1}\int^1_\alpha dx
\frac{(1-x)}{(1-\al)}\biggl(\frac{1}{x}\biggr)\biggl[\cdots\biggr]
\nonumber\\
&=&\lim_{\alpha\to 1}\int^1_0 dz
\frac{(1-\alpha)(1-z)}{\alpha + (1-\alpha)z}\biggl[\cdots\biggr],
\eea
where the variable change $x=\alpha + (1-\alpha)z$ was made and the
terms in $\biggl[\cdots\biggr]$ are regular in the $\al \rightarrow 1$
limit. Thus, $I^{\rm zm}_A$ vanishes in the $\alpha\to 1$ limit. Note
that the factor $1/x$ in Eq.~(\ref{fz:2}) comes from $S^{+(h=0)}_{\rm
inst}$ and $(1-x)/(1-\alpha)$ from the energy denomenator combined with
the prefactor in Eq.~(\ref{fz:1}).

Therefore, we conclude that the form factor $f(q^2)$ is immune to the
zero-mode contrary to the discussion made by Jaus~\cite{Jaus,Jaus02},
where a zero-mode contamination in the form factor $f(q^2)$ was
claimed. As we discussed in Section \ref{sect.I}, our manifestly
covariant formulation should be distinguished from the formulation
involving a light-like four-vector $\omega (\omega^2 = 0)$.  This is
one of the main observations in our present work.

For the readers who are interested in checking our numerical results
for the form factors in the $q^+=0$ frame, we present in Appendix C the
exact LF valence expressions (equivalent to the covariant result) for
the form factor $f(q^2)$ as well as $g(q^2)$ and $a_+(q^2)$ that are
obtained by the Feynman parametrization in the $q^+=0$ frame.

In the following subsection, \ref{SecIIIE}, we check if the absence of the
zero-mode in $f(q^2)$ is still valid in the case of the vector meson
vertex used frequently in the light-front quark model 
(LFQM) calculations.

\subsection{Vector meson vertex in LFQM}
\label{SecIIIE}
A vector meson vertex frequently used in 
LFQM calculations~\cite{Jaus90,Jaus96,CCH,Jaus,Jaus02,HC} is given by
\begin{equation}{\label{RV:1}}
 \Gamma^\mu = \gamma^\mu - \frac{(p_2-k)^\mu}{M'_{0}+m_2 + m}.  
\end{equation} 
This vertex is denoted by $\Gamma^\mu_{\rm LFQM}$ in the remainder of this
paper.  We check in this subsection whether substitution of this form
of $\Gamma^\mu$ in Eq.~(\ref{eqq:4}) instead of the simple vertex
$\Gamma^\mu = \gamma^\mu$ would affect our finding in the previous
subsection, {\it i.e.}, the absence of a zero-mode in $f(q^2)$.

Denoting the trace for the second term in Eq.~(\ref{RV:1}) by
$T^+_h$(see $S^+_h$ in Eq.~(\ref{ch:0}) for the first term), 
we obtain
\bea{\label{RV:2}}
&&\hspace{-0.5cm}T^{+}_{h}
= T^{+(h)}_{V}-T^{+(h)}_A
\nonumber\\
&&=-4\frac{(p_2-k)\cdot\epsilon^*(h)}{M'_0+m_2+m}
\biggl[
i\epsilon^{+\mu\nu\sigma}(p_{1\rm on})_\mu(p_{2\rm on})_\nu(k_{\rm on})_\sigma
\nonumber\\
&&+ (p_{2\rm on}\cdot k_{\rm on}-m_2m)p^+_1
+(p_{1\rm on}\cdot k_{\rm on}+m_1m)p^+_2
\nonumber\\
&&- (p_{1\rm on}\cdot p_{2\rm on}+m_1m_2)k^+
+ (k^--k^-_{\rm on})p^+_{1\rm on}p^+_{2\rm on}\biggr],
\nonumber\\
\eea
for the plus current matrix element. Note that the first term ({\it
i.e.}, the term including $\epsilon^{+\mu\nu\sigma}$) in
Eq.~(\ref{RV:2}) corresponds to the vector current and the rest
to the axial-vector current contribution.  We use Eq.~(\ref{sep}) to 
obtain
the last term, $(k^--k^-_{\rm on})p^+_{1\rm on}p^+_{2\rm on}$, which
vanishes in the valence diagram.  We do not separate the on-shell
propagating part from the instantaneous one in $T^+_h$ as we did in
$S^+_h$ due to the complication of the form arising from the
$(p_2-k)$-term in Eq.~(\ref{RV:2}).

The total trace $({\cal T}^+_h)_{\rm LFQM}$ for 
the vertex $\Gamma^+_{\rm LFQM}$ is then given by
\bea\label{RV:3}
({\cal T}^+_h)_{\rm LFQM} = S^+_h - T^+_h.
\eea
The complete expressions
for the form factors with the vertex $\Gamma^\mu_{\rm LFQM}$
are presented in Appendix D.

Because the only suspected term for the zero-mode contribution is
$T^{+(h=0)}_{A}$ in Eq.~(\ref{RV:2}), we shall discuss whether this
term gives a nonvanishing zero-mode contribution to the weak form
factor $f(q^2)$ in the $q^+=0$ limit.

To investigate the zero-mode contribution from $T^{+(h=0)}_A$,
we use the same argument discussed in the previous subsection, but
replacing $S^{+(h=0)}_{\rm inst}(k^-=k^-_{\Lambda_1}\,{\rm or}\,
k^-_{m_1})$ with $T^{+(h=0)}_A(k^-=k^-_{\Lambda_1}\,{\rm
or}\,k^-_{m_1}) \equiv [T^{+(h=0)}_A]_{\rm zm}$. The explicit form of
$[T^{+(h=0)}_A]_{\rm zm}$ is given by Eq.~(\ref{eq:D7}) in Appendix D.

Showing again only the longitudinal momentum fraction factors relevant
to the zero-mode from Eq.~(\ref{eq:D7}) in Appendix D, we find the 
nonvanishing term in the limit of $\alpha\to 1$ (or equivalently $x\to 
1$) as 
\bea{\label{RV:4}}
[T^{+(h=0)}_A]_{\rm zm}\sim \sqrt{\frac{1}{1-x}}\biggl[\cdots\biggr],
\eea
where the factor $\biggl[\cdots\biggr]$ corresponds to the regular
part.  Equation~(\ref{RV:4}) holds both for the  $k^-=k^-_{m_1}$ and
$k^-_{\Lambda_1}$ cases. However, it is very interesting to note that
even though $[T^{+(h=0)}_A]_{\rm zm}$ in Eq.~(\ref{RV:4}) itself shows
singular behavior as $x\to 1$, the net result of the zero-mode
contribution is given by
\bea\label{RV:5}
I^{\rm zm}_A &\sim& \lim_{\alpha\to1}\int^1_\alpha dx
\frac{(1-x)}{(1-\alpha)}\sqrt{\frac{1}{1-x}}\biggl[\cdots\biggr]
\nonumber\\
&=&\lim_{\alpha\to 1}\int^1_0 dz
\frac{(1-\alpha)(1-z)}{\sqrt{(1-\alpha)(1-z)}}\biggl[\cdots\biggr]
\eea
where the factor $\biggl[\cdots\biggr]$ again corresponds to the regular 
part. Thus, $I^{\rm zm}_A$ vanishes as $\alpha\to 1$ and
our conclusion for the vanishing zero-mode contribution
to the form factor $f(q^2)$ in the $q^+=0$ frame holds even in the 
vector meson LF vertex $\Gamma^\mu_{\rm LFQM}$,
which is frequently used for the more realistic LFQM analysis. 

\section{Numerical Results}
{\label{sect.IV}}

In this section, we present the numerical results for the transition
form factors and verify that all of the four form factors
($g(q^2),a_\pm(q^2),f(q^2)$) obtained in the LF formulation are in
complete agreement with the manifestly covariant results presented in
Section \ref{sect.II}. We also confirm that the numerical results of
$g(q^2)$, $a_+(q^2)$, and $f(q^2)$ obtained in the $q^+ = 0$ frame are
identical to those obtained in the purely longitudinal $q^+ > 0$ frame,
as they should be. We do not aim at finding the best-fit parameters to
describe the experimental data in this work.  As we mentioned earlier,
however, our model calculations have a generic structure and the
essential findings from our calculations are expected to apply to the
more realistic models, although the quantitative results would differ
from each other depending on the details of the dynamics in each
model.

The used model parameters for $B$, $D^*$, and $\rho$ mesons are 
$M_B = 5.28$ GeV, $M_{D^*} = 2.01$ GeV, $M_\rho = 0.771$ GeV,
$m_b=4.9$ GeV, $m_c=1.6$ GeV, $\Lambda_b=10$ GeV, $\Lambda_c=5$ GeV,
$g_B=5.20$, and $g_D^*=3.23$,
as well as $m_u=m_d=0.43$ GeV, $\Lambda_u=1.5$ GeV, and $g_\rho=5.13$.

These parameters are fixed from the nomalization conditions of the
pseudoscalar and vector meson elastic form factors at $q^2 =0$.  The
manifestly covariant results for these form factors and also the decay
constants are summarized in Appendix B. The decay constants (see
Eqs.~(\ref{eq:B6}) and~(\ref{eq:B13})) of $B$ and $D^*$ obtained from
the above fixed parameters are $f_\rho = 274$ MeV, $f_{D^*}=216$ MeV
and $f_B=150$ MeV, which are within the range used in
Refs.~\cite{CCH,Jaus,BMNS,Bernard,CJBK}.
\begin{figure*}
\psfig{figure=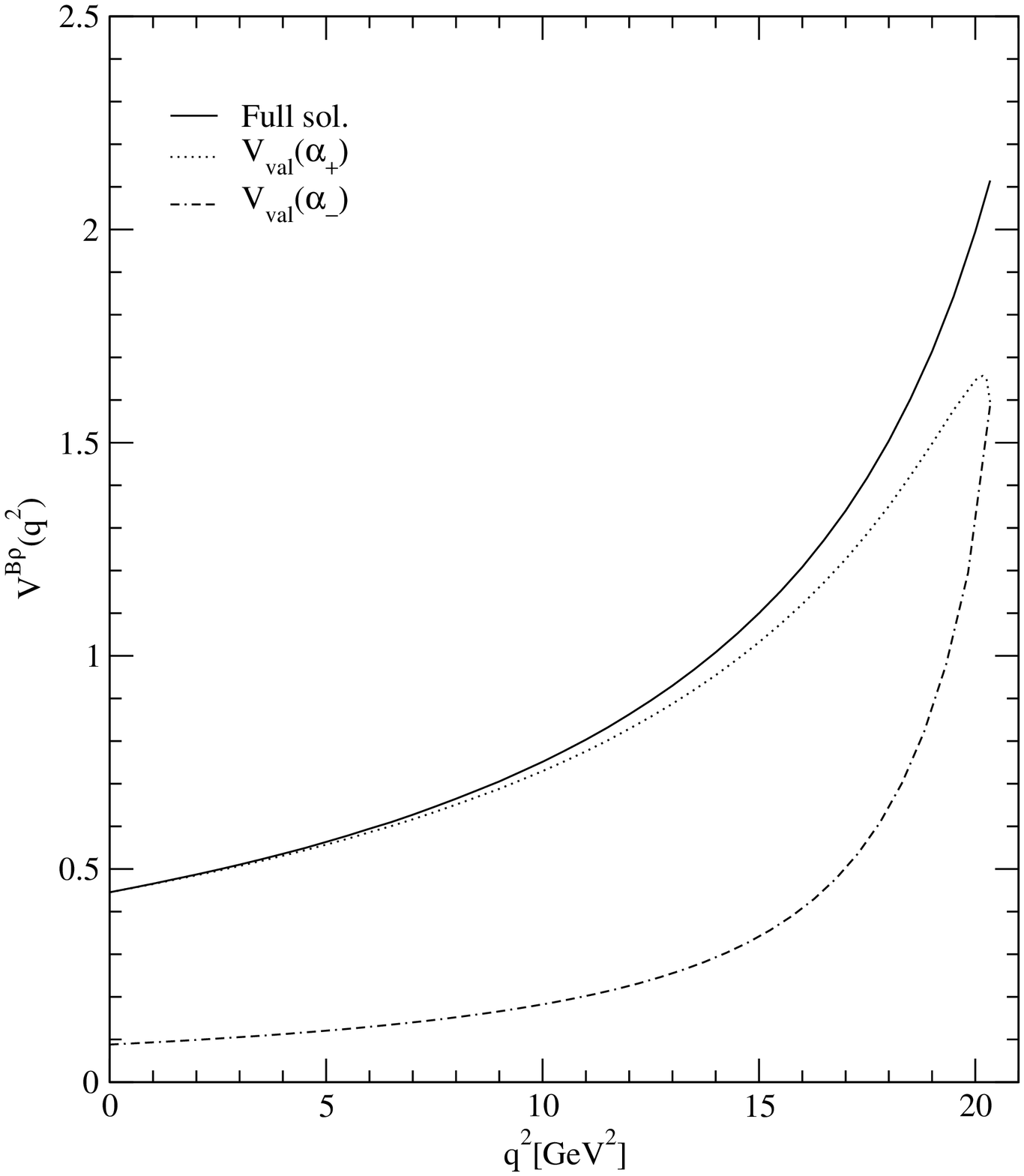,height=10cm,width=8cm}
\psfig{figure=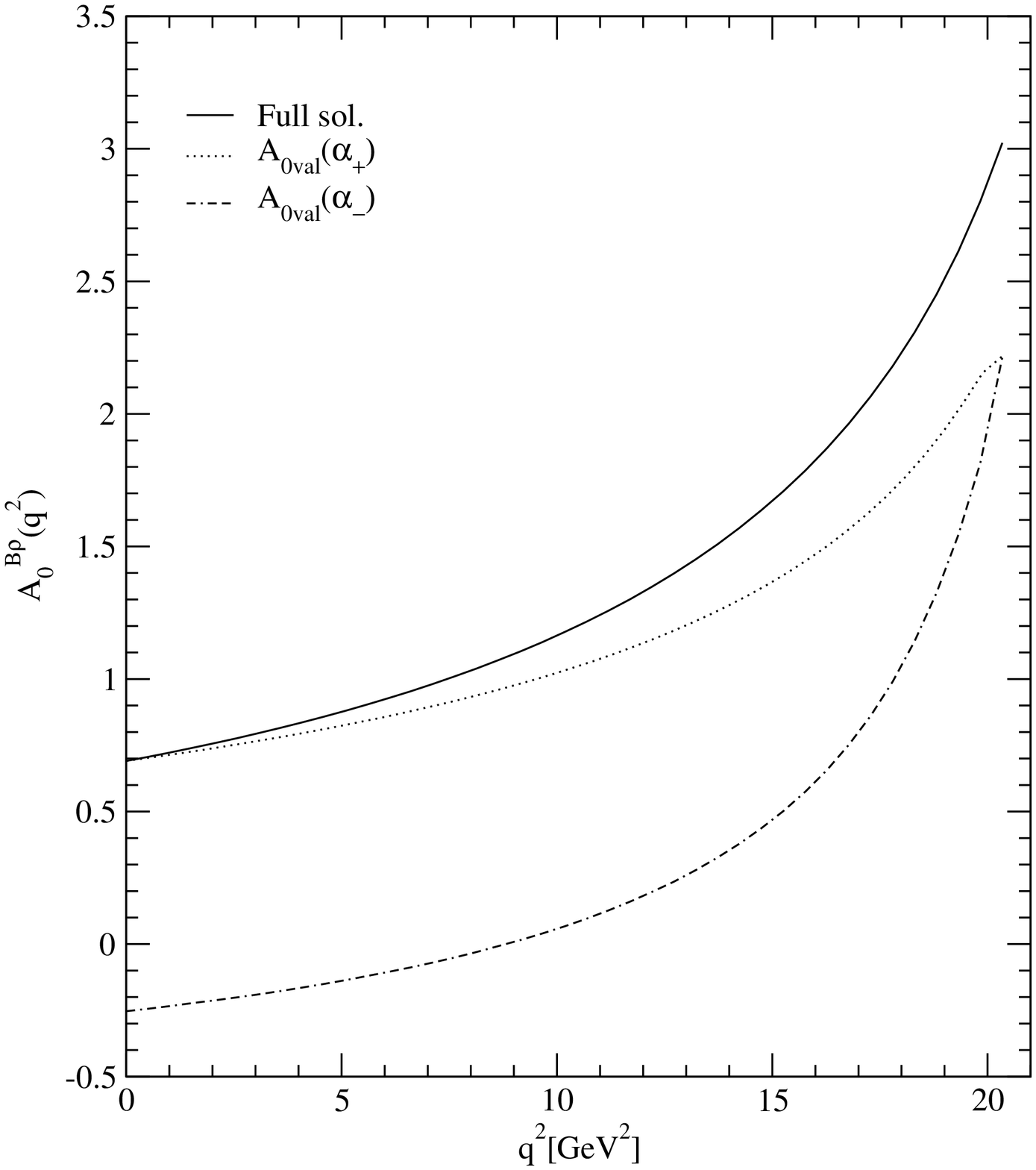,height=10cm,width=8cm}
\psfig{figure=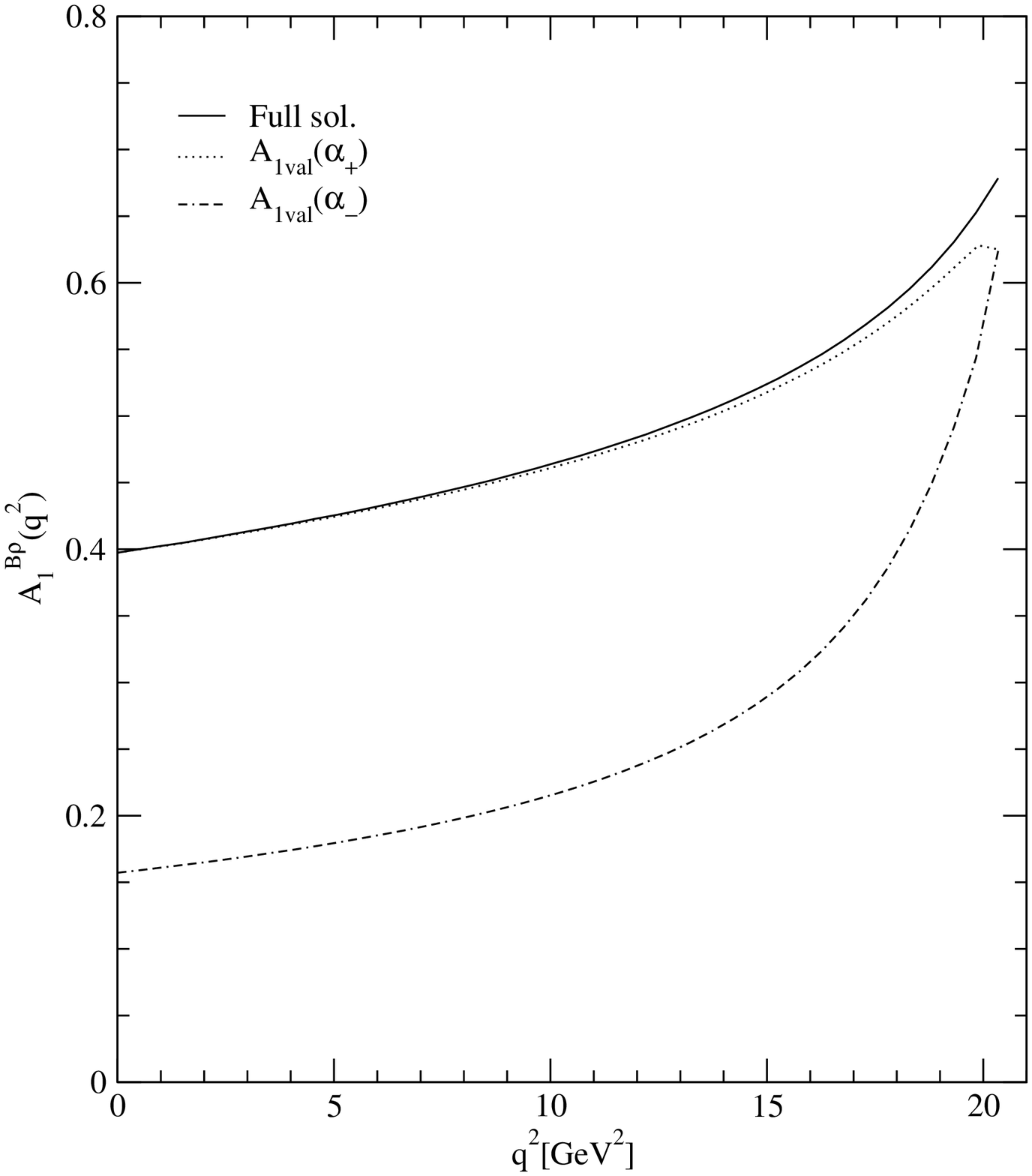,height=10cm,width=8cm}
\psfig{figure=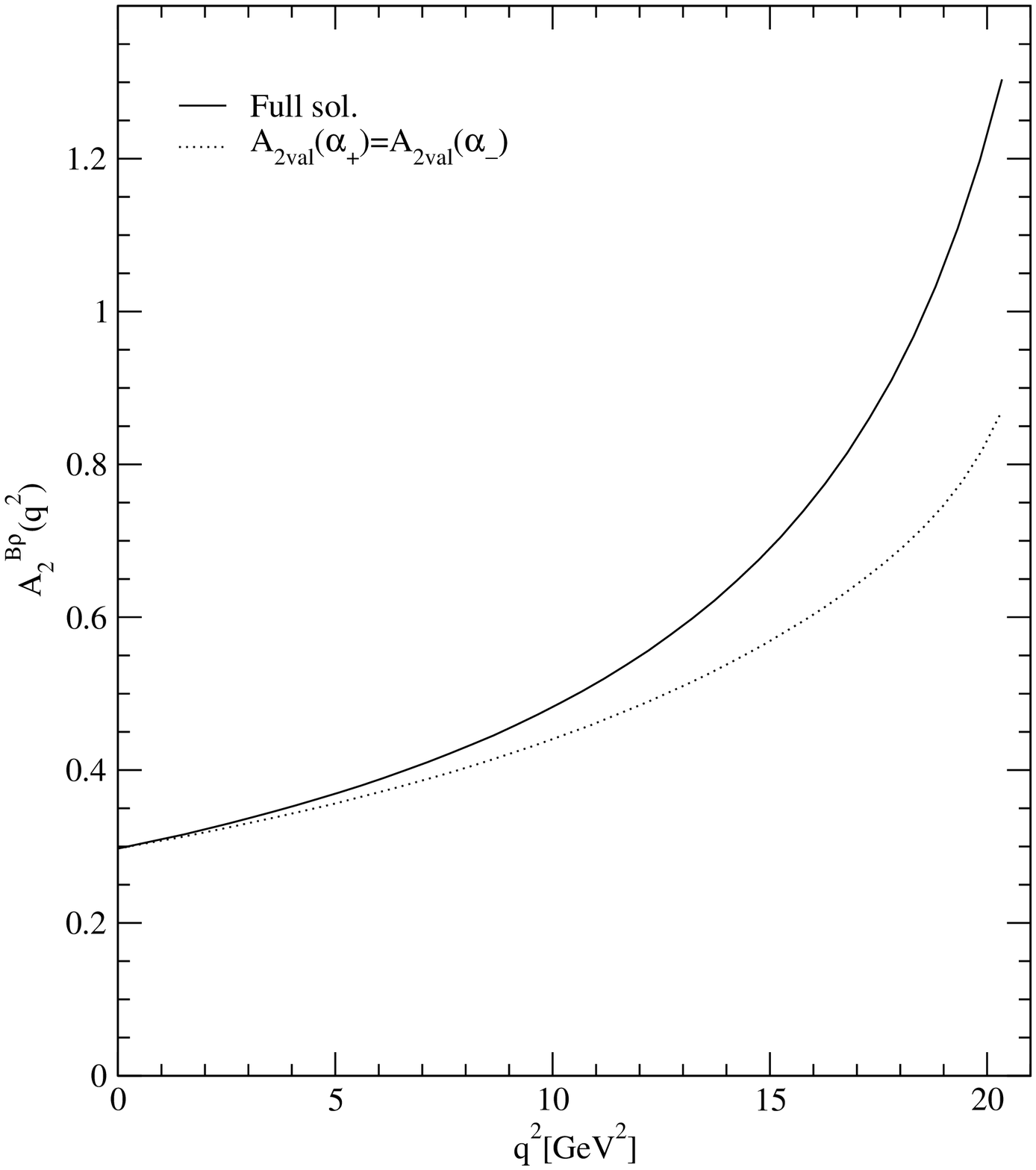,height=10cm,width=8cm}
\caption{Weak form factors for the $B\to \rho$ transition obtained
from the purely longitudinal frame. The solid, dotted, and
dot-dashed lines represent the full (val + nv) solution, the valence
contribution with $\alpha_+$-dependence, and the valence contribution
with $\alpha_-$-dependence, respectively. The full solution is
exactly identical to the covariant one.  {\label{Fig_Brho}}}
\end{figure*}

In Fig.~\ref{Fig_Brho}, we present the weak form factors defined in
Eq.~(\ref{eq:2}) for the $B \to \rho$ (heavy-to-light) transition.
Since the weak form factors $V,A_1$, and $A_2$ do not involve $a_-$, we
computed these form factors both in the $q^+ = 0$ DYW frame and in the
purely longitudinal $q^+ > 0$ frame. The full results depicted by the
solid lines are in complete agreement regardless of the choice of
frames, as they should be. In the $q^+ >0$ frame, we can separate the
full result into the valence contribution and the nonvalence
contribution. To show this, we present the valence contribution
computed in the two recoil directions given by Eqs.~(\ref{pl:2}) and
(\ref{apm}), {\it i.e.}, $\alpha_+$(dotted line) and
$\alpha_-$(dot-dashed line). 
Note that $a_+$ and $a_-$ are obtained using both $\alpha_+$ and 
$\alpha_-$ solutions as shown in Eq.(\ref{pl:4}) and thus $A_2(q^2)$
in Fig.\ref{Fig_Brho} doesn't have any distinction in the valence 
contributions between $A_{2val}(\alpha_+)$ and $A_{2val}(\alpha_-)$.
Of course, the nonvalence contributions
are obtained by subtracting the valence contributions from the full
results. We have also confirmed the agreement of the full results
(solid lines) and the manifestly covariant results presented in Section
\ref{sect.II}.

\begin{figure*} 
\psfig{figure=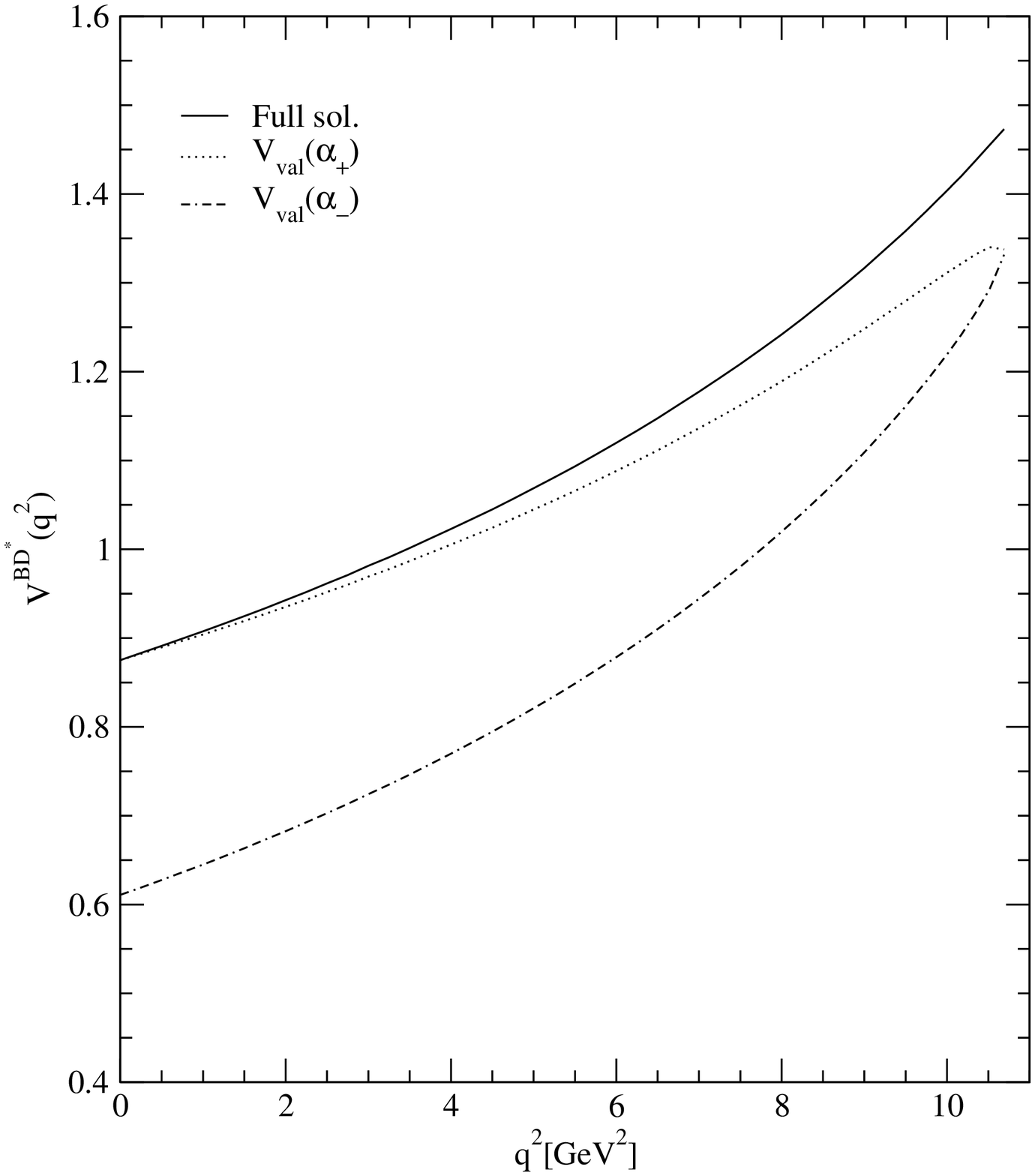,height=10cm,width=8cm}
\psfig{figure=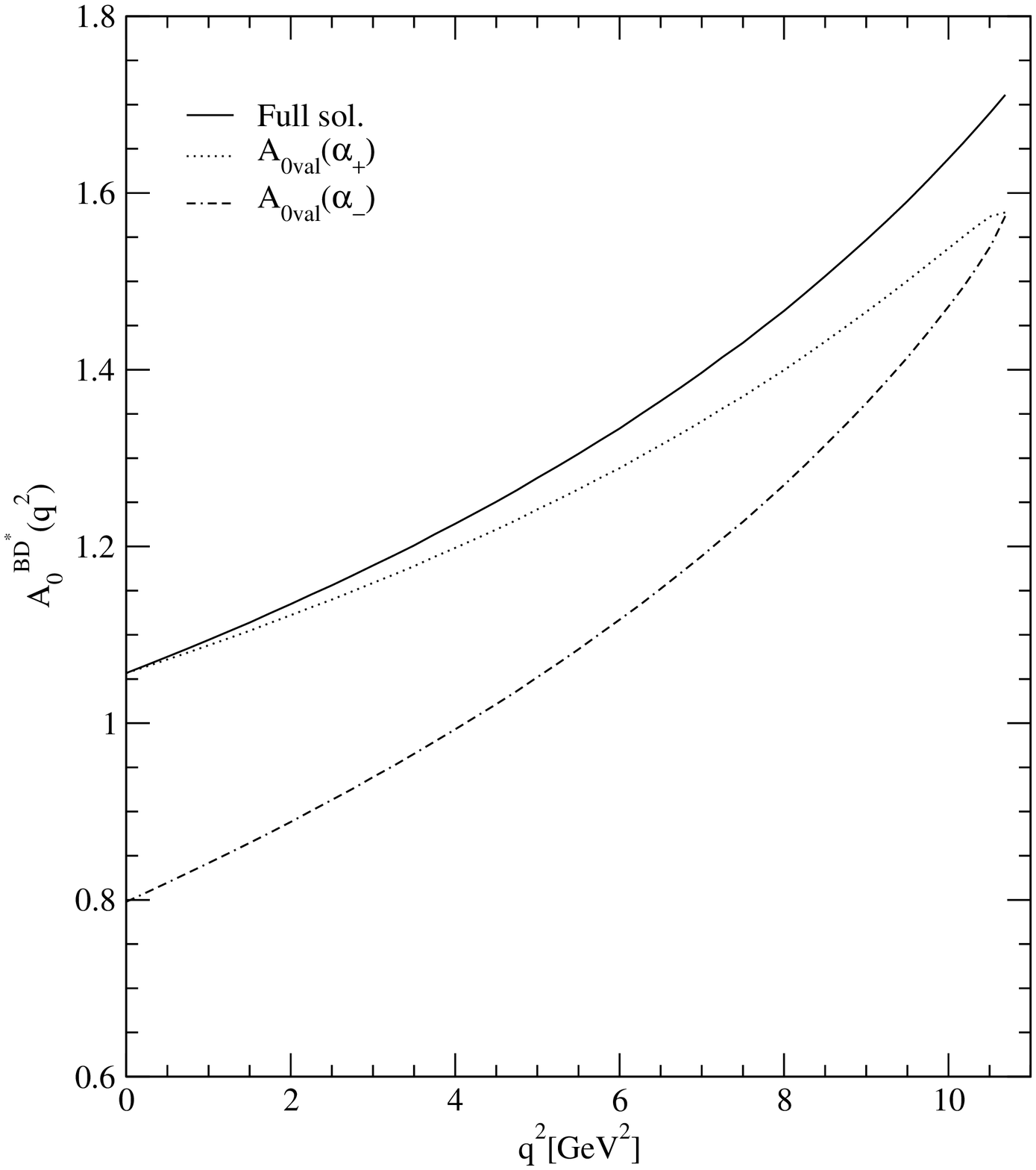,height=10cm,width=8cm}
\psfig{figure=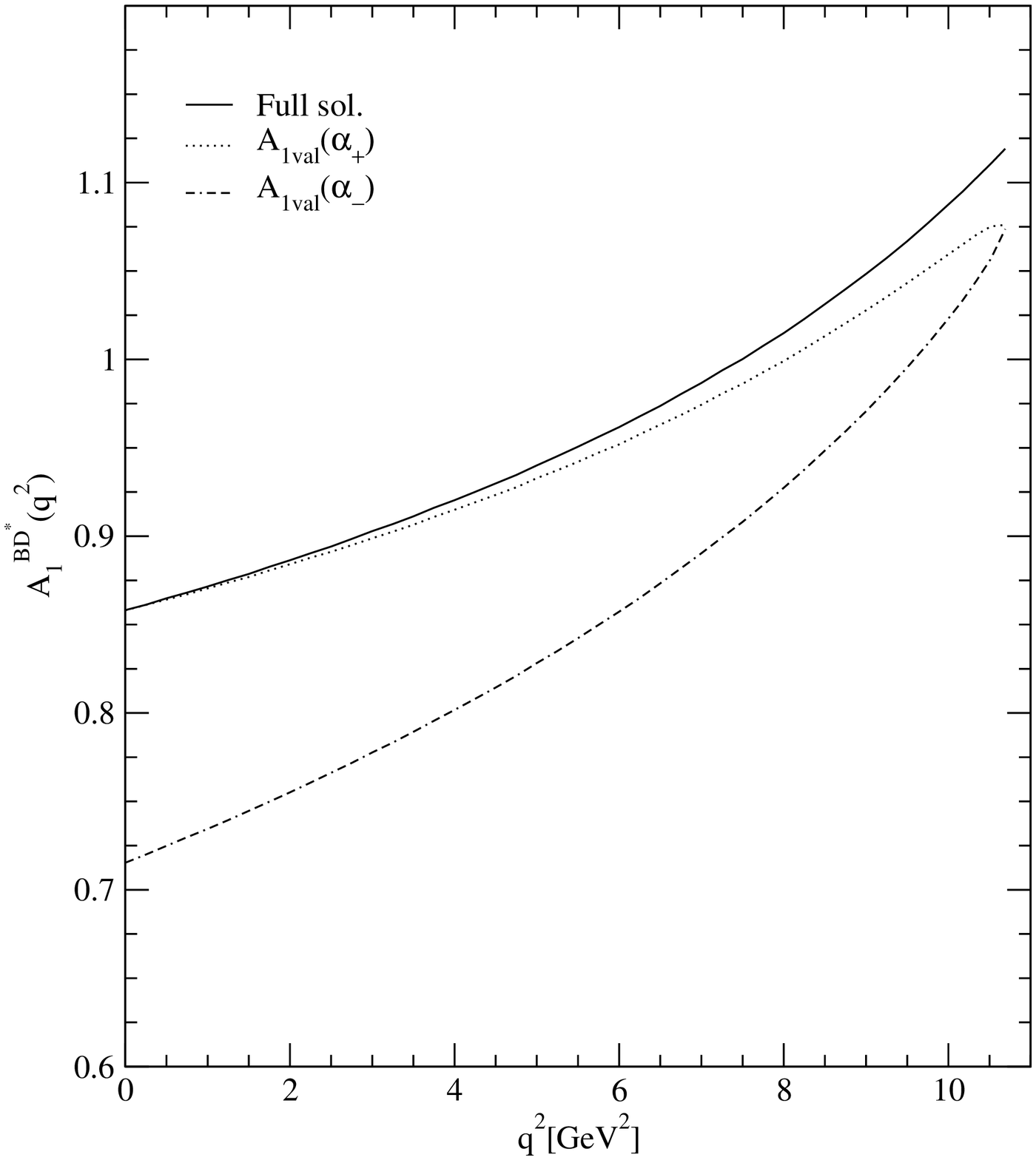,height=10cm,width=8cm}
\psfig{figure=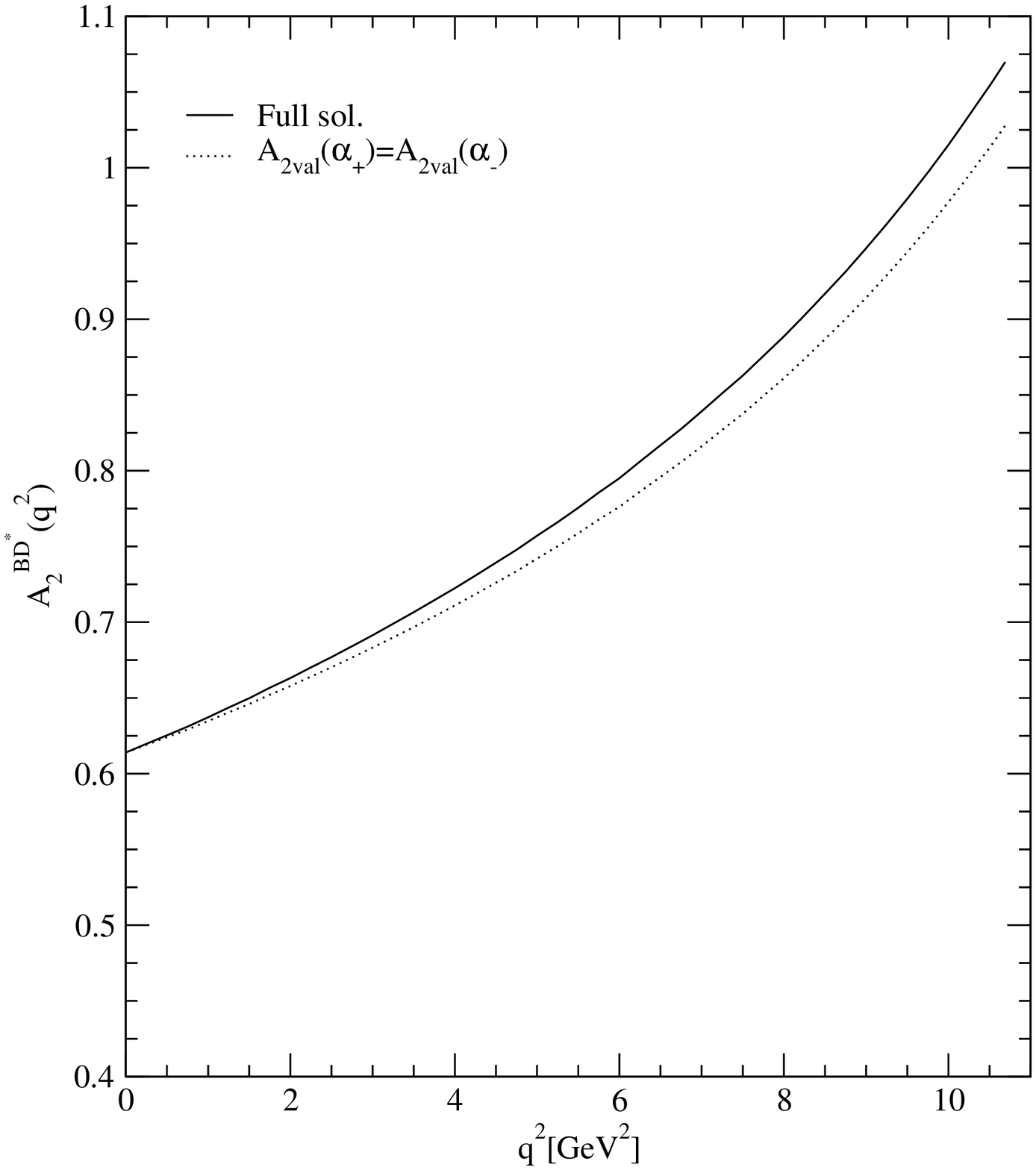,height=10cm,width=8cm} 
\caption{Weak form factors for the $B\to D^*$ transition obtained from
the purely longitudinal frame. The solid, dotted, and dot-dashed lines
represent the full (val + nv) solution, the valence contribution with
$\alpha_+$-dependence, and the valence contribution with
$\alpha_-$-dependence, respectively. The full solution is exactly the
same as the covariant one.  {\label{Fig_BD}}}
\end{figure*}

In Fig.~\ref{Fig_BD}, we present the same for the $B \to D^*$
(heavy-to-heavy) transition. The general features are similar to the
case of the heavy-to-light meson decay shown in Fig.~\ref{Fig_Brho}.
However, one can see that the nonvalence contributions are
significantly reduced in the heavy-to-heavy case. 
Experimentally, two form-factor ratios for $B\to D^*$ decays defined
by~\cite{Neu,CLEO}
\bea\label{R1R2}
R_1(q^2)&=&\biggl[1-\frac{q^2}{(M_B + M_{D^*})^2}\biggr]
\frac{V(q^2)}{A_1(q^2)},\nonumber\\
R_2(q^2)&=&\biggl[1-\frac{q^2}{(M_B + M_{D^*})^2}\biggr]
\frac{A_2(q^2)}{A_1(q^2)},
\eea
have been measured by CLEO~\cite{CLEO} as $R_1(q^2_{\rm max})=1.24\pm
0.26\pm 0.12$ and $R_2(q^2_{\rm max})=0.72\pm 0.18\pm 0.07$.  We obtain
$R_1(q^2_{\rm max})=1.05$ and $R_2(q^2_{\rm max})=0.76$, which are
compatible with the these data and other theoretical predictions:
$R_1(q^2_{\rm max})=1.35$ and $R_2(q^2_{\rm max})=0.79$ in
Ref.~\cite{Neu}, $R_1(q^2_{\rm max})=1.27$ and $R_2(q^2_{\rm
max})=1.01$ in Ref.~\cite{ISGW2}, and $R_1(q^2_{\rm max})=1.24$ and
$R_2(q^2_{\rm max})=0.91$ in Ref.~\cite{AW}.

The form factor $a_-(q^2)$ was also constrained by 
the flavor independence in Ref.~\cite{AW} as  
\bea\label{hth}
a_+ (q^2_{\rm max}) -a_-(q^2_{\rm max}) & = &-\frac{1}{\sqrt{M_{D^*}M_B}}. 
\eea 
Our value, $a_+ - a_- \sim-0.36$ at $q^2_{\rm max}$, is consistent 
with Eq.~(\ref{hth}) which yields $a_+-a_-\sim-0.31$. 
The form factor $a_+(q^2)$ was further constrained by the
flavor independence in the heavy quark limit~\cite{AW} and given by 
\bea\label{hthp}
a_+ (q^2_{\rm max}) &=&-\frac{1}{\sqrt{4M_{D^*}M_B}}\biggl[ 1 + 
\frac{M_{D^*}}{M_B}\biggl(1-\frac{M_{D^*}}{m_c}\biggr)\biggr].
 \nonumber\\
\eea 
This yields the value $a_+(q^2_{\rm max}) \sim -0.14$, which is very close to 
our value $a_+ \sim -0.15$. 

\begin{table}
\caption{The calculated $B\to \rho$ transition form factors at
$q^2=0$.{\label{tab.1}}}
\begin{ruledtabular}
\begin{tabular}{lrrrr}
Ref. & V & $A_0$ & $A_1$ & $A_2$  \\
\hline
This work & 0.45 &0.69 & 0.39 & 0.30 \\
\hline
LCSR\protect\cite{Ball} & $0.6(2)$ &--&$0.5(1)$ &
$0.4(2)$ \\
\hline
LAT\protect\cite{Deb} &$0.35^{+0.06}_{-0.05}$
& $0.30^{+0.06}_{-0.04}$ & $0.27^{+0.05}_{-0.04}$
& $0.26^{+0.05}_{-0.03}$ \\
\hline
QM\protect\cite{Jaus96} & 0.35 & -- & 0.26 & 0.24\\
\end{tabular}
\end{ruledtabular}
\end{table}

\begin{table}
\caption{The calculated $B\to D^*$ transition form factors at
$q^2=0$.{\label{tab.2}}}
\begin{ruledtabular}
\begin{tabular}{lrrrr}
Ref. & V & $A_0$ & $A_1$ & $A_2$  \\
\hline
This work & 0.89 &1.07 & 0.87 & 0.62  \\
\hline
QM\protect\cite{Jaus96} &0.81 &- & 0.69 & 0.64 \\
\hline
QM\protect\cite{MS} &0.76 &0.69 & 0.66 & 0.62 \\
\end{tabular}
\end{ruledtabular}
\end{table}

Our results for the $B \to \rho$ and $B \to D^*$ transition form
factors at $q^2=0$ are also compared with other theoretical results 
in Tables \ref{tab.1} and \ref{tab.2},
respectively.

In the following subsection, we present the frame dependence of the 
individual valence and nonvalence contributions using the typical frames
summarized in Appendix A.

\subsection{Frame dependence}
\label{sec.IVA}
\begin{figure}[t]

\vspace{1ex}

\psfig{figure=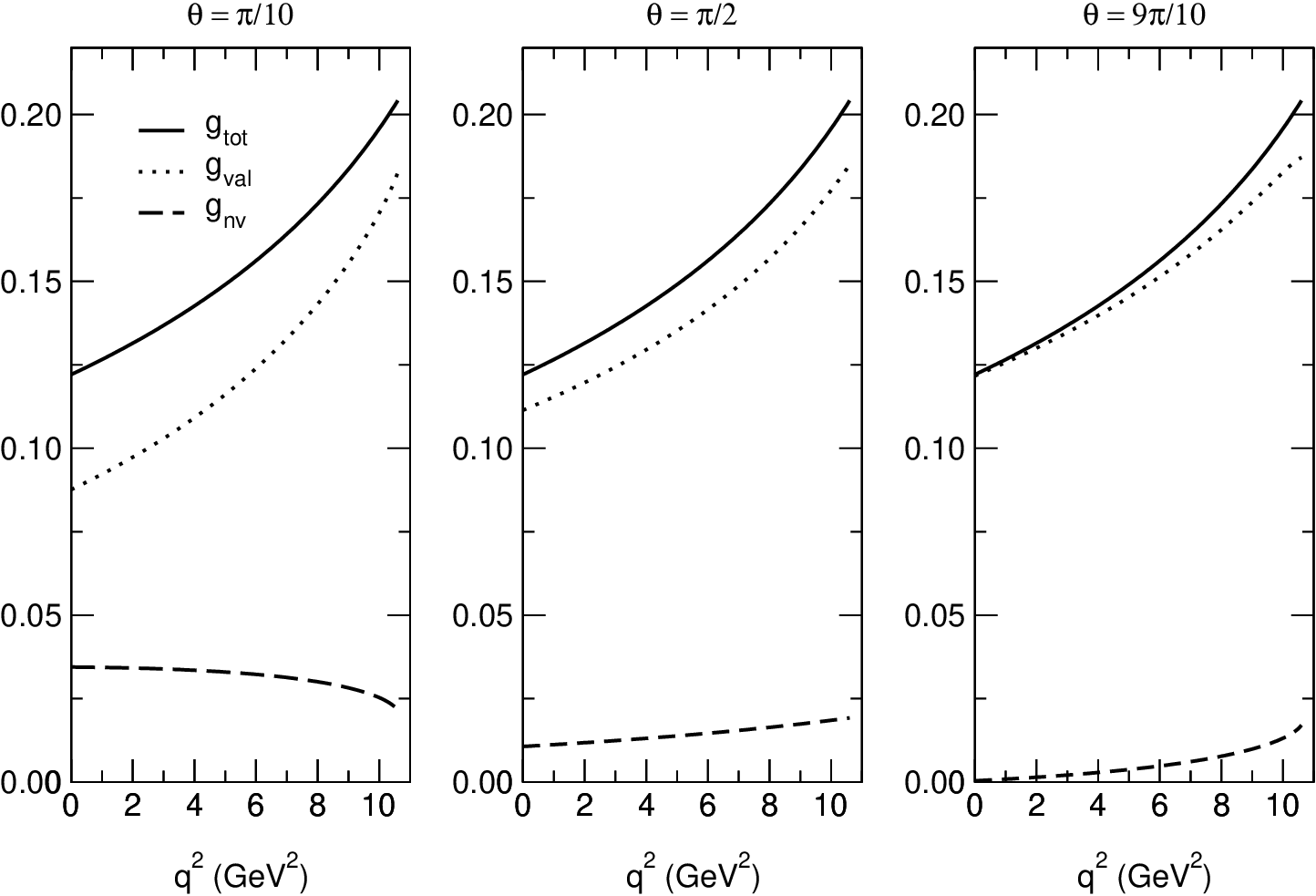,height=5.4cm,width=8cm}

\vspace{2ex}

\caption{Breit frame $g$ form factor for $B \to D^*$. \label{fig.BRTg_D}}
\end{figure}

\begin{figure}[t]

\vspace{5ex}

\psfig{figure=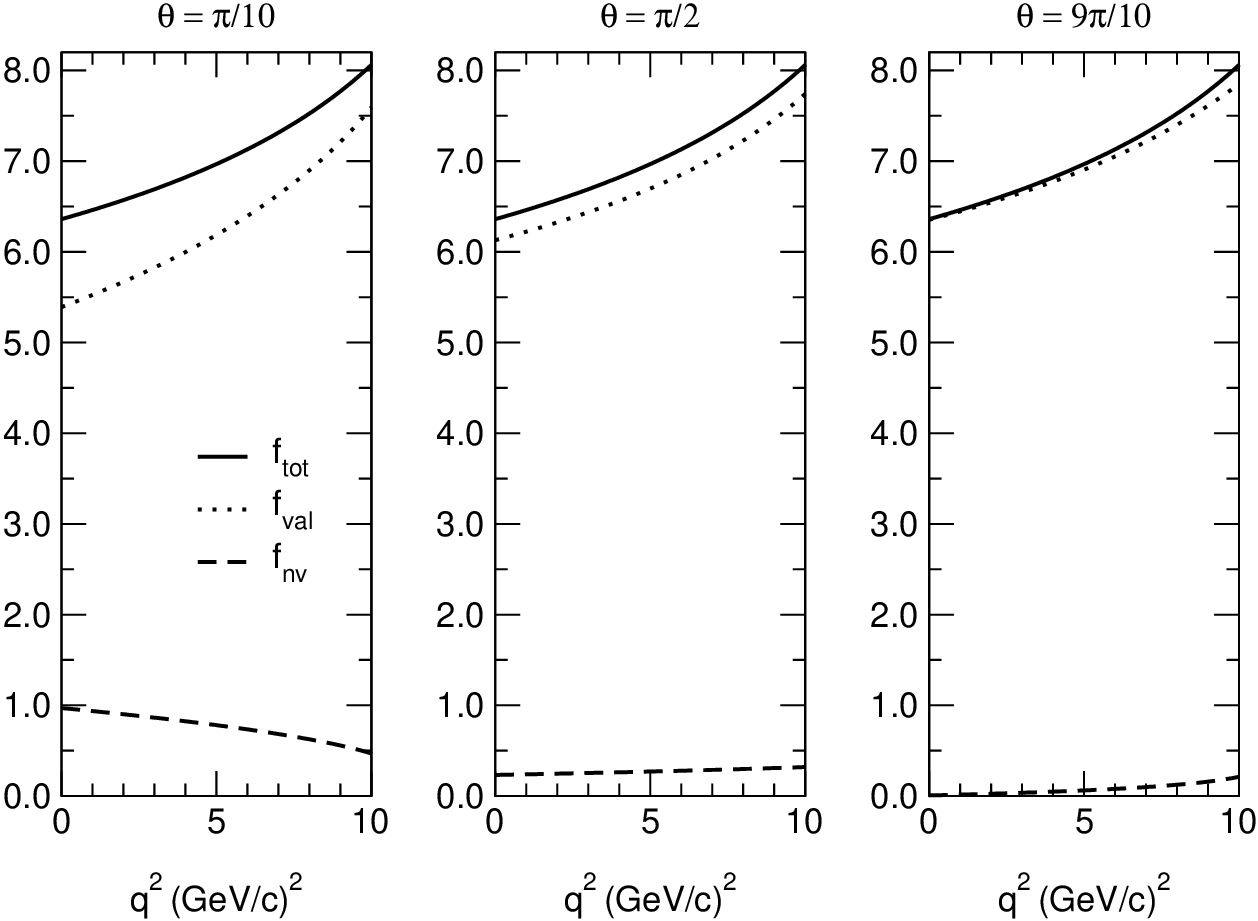,height=5.4cm,width=8cm}

\vspace{2ex}

\caption{Breit frame $f$ form factor for $B \to D^*$. \label{fig.BRTf_D}}
\end{figure}

We show the frame dependence of the form factors $g$ and $f$ for $B \to
D^*$. In Figs.  \ref{fig.BRTg_D} and \ref{fig.BRTf_D} we plotted these
form factors in the Breit frame for three different orientations of the
momentum transfer.  The general trend we see is that the contribution
to the form factor from the nonvalence diagram becomes smaller as the
angle $\theta$ increases. 
For $\theta = \pi$, note that $q^+=0$ at $q^2=0$. Thus, the suppression
of the nonvalence contribution for larger angles,  close to $\theta = \pi$,
is natural especially in the region near $q^2 = 0$.
We found little difference between the
results calculated in the Breit frame with the ones calculated in the
target-rest frame, so we do not plot the latter ones.  

\begin{figure}[t]

\vspace{3ex}

\psfig{figure=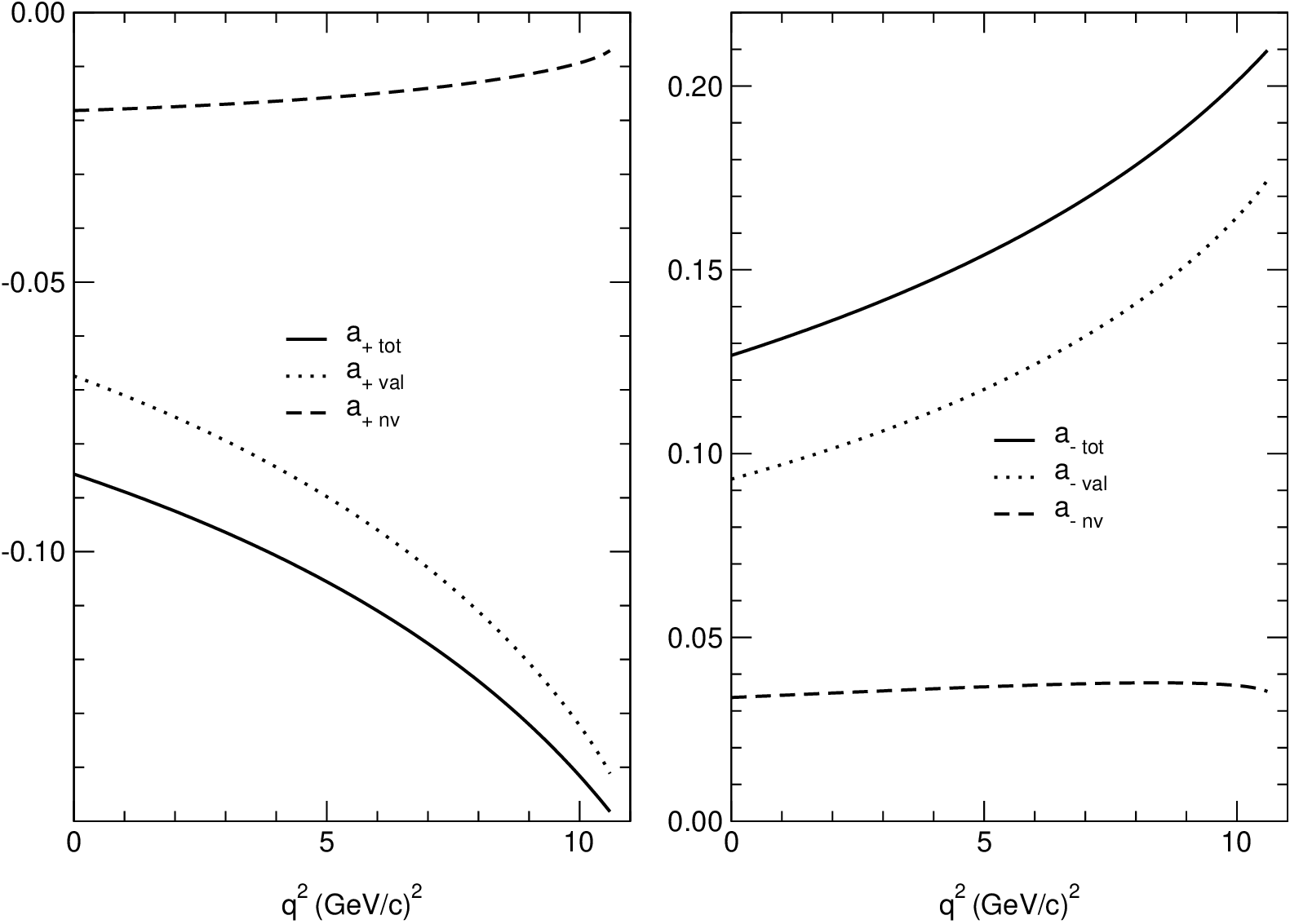,height=5cm,width=8cm}
\caption{Breit frame $a_\pm$ form factors for $B \to D^*$.\label{fig.BRTa_D}}
\end{figure}
\begin{figure}[t]

\vspace{1ex}

\psfig{figure=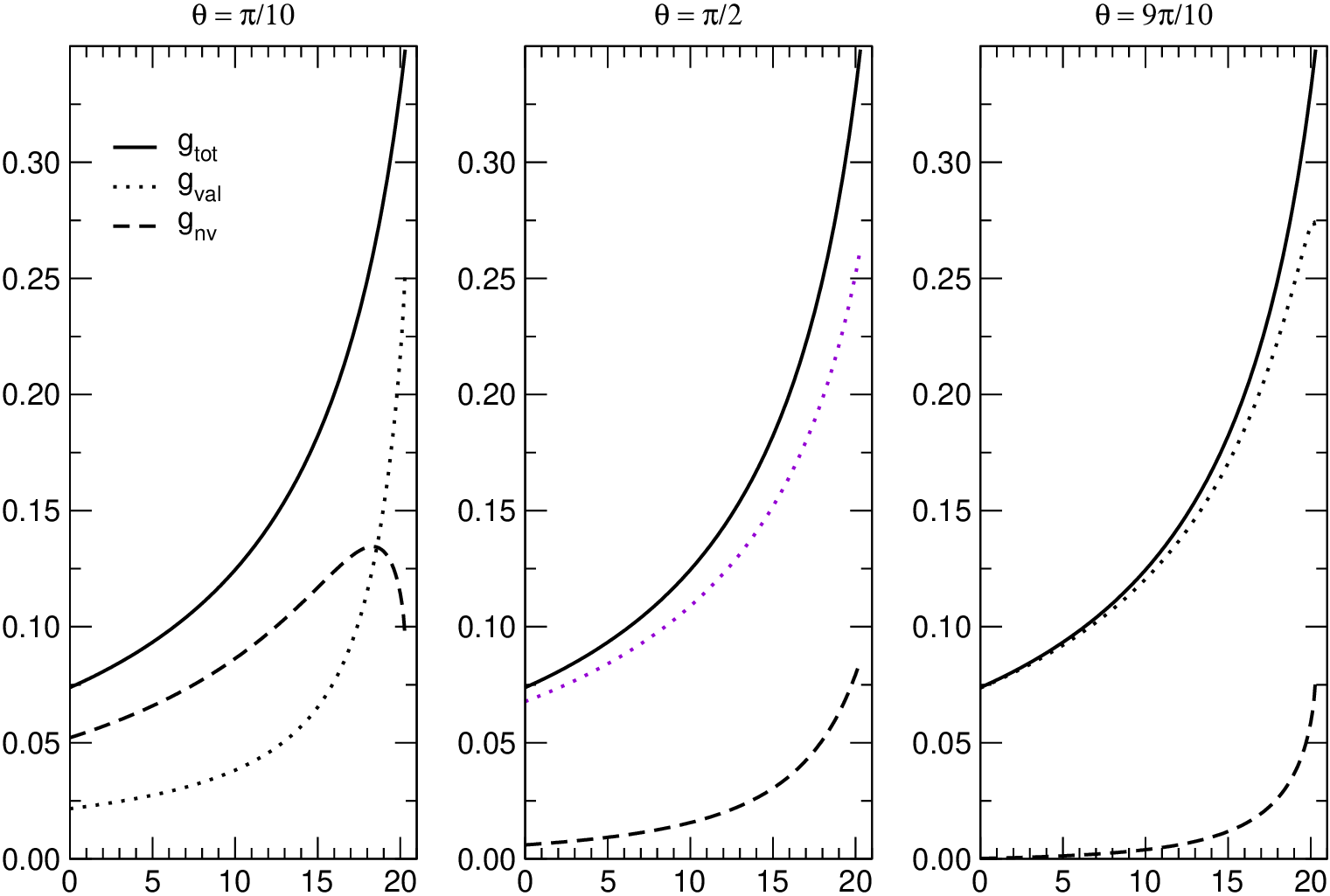,height=5cm,width=8cm}

\vspace{2ex}

\caption{Breit frame $g$ form factor for $B \to \rho$. \label{fig.BRTg_rho}}
\end{figure}
\begin{figure}[t]

\vspace{1ex}

\psfig{figure=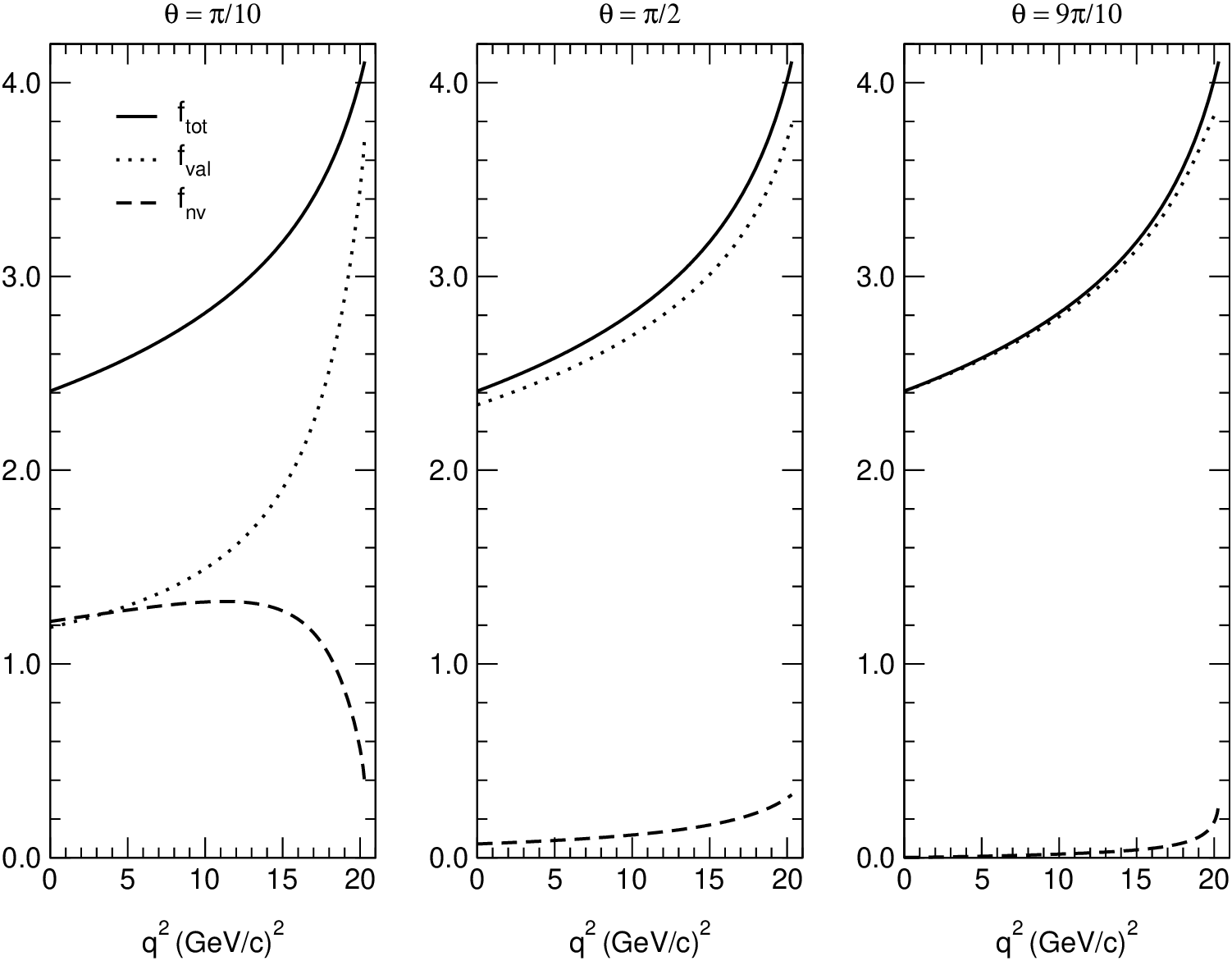,height=5cm,width=8cm}

\vspace{2ex}

\caption{Breit frame $f$ form factor for $B \to \rho$.\label{fig.BRTf_rho}}
\end{figure}
\begin{figure}[t]
\psfig{figure=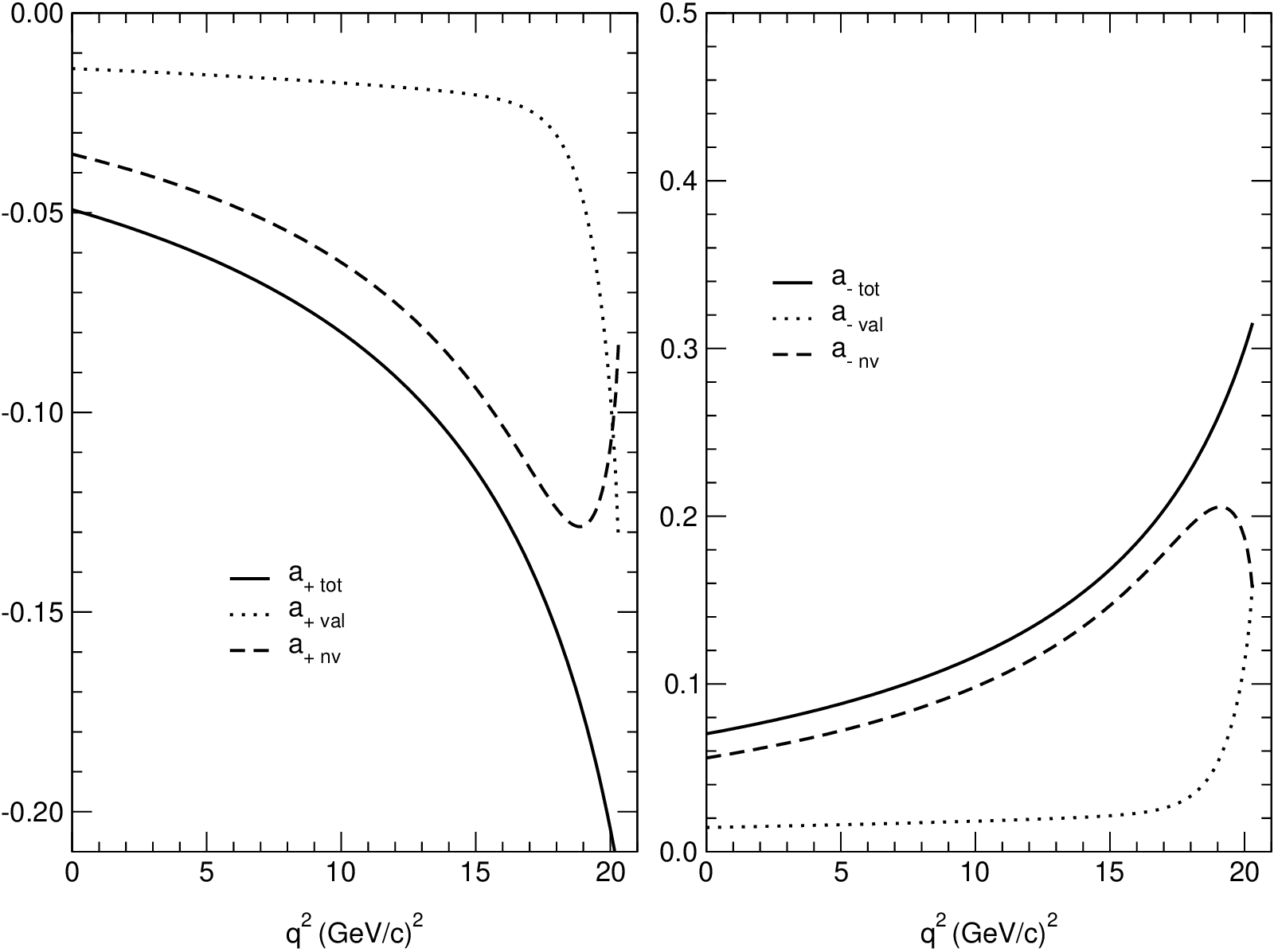,height=5cm,width=8cm}
\caption{Breit frame $a_\pm$ form factors for $B \to \rho$. \label{fig.BRTa_rho}}
\end{figure}
We show the form factors $a_\pm$ in the Breit frame for $B \to D^*$ in
Fig.~\ref{fig.BRTa_D}. As explained before, we can only extract these
form factors if we combine the calculations for two values of the polar
angle $\theta$, {\it i.e.}, two values for $\alpha$.  Therefore, we do
not plot the results for different values of $\theta$.
In Fig.\ref{fig.BRTa_D}, the used values of the polar angle are $\theta = 
\pi/10$  and $9\pi/10$.

The results for the heavy-to-light decay $B \to \rho$ are given in
Figs.~\ref{fig.BRTg_rho}-\ref{fig.BRTa_rho}.  The qualitative
difference between the heavy-to-heavy and the heavy-to-light decay
mentioned before is clearly seen in these figures too. The nonvalence
parts become more prominent for the heavy-to-light case. In both cases
the nonvalence contributions to $g$ and $f$ are suppressed for
increasing polar angle $\theta$.

\section{Conclusion}
{\label{sect.V}}

In this work, we analyzed the transition form factors between
pseudoscalar and vector mesons using both the manifestly covariant
calculation and the light-front calculation for $\la J_{V-A}^+ \ra$. In
LFD, we presented three results: one from the DYW ($q^+ =0$) frame, the
other from the purely longitudinal $q^+ > 0$ frame, and finally results
obtained in the Breit frame.  In the DYW ($q^+ = 0$) frame, the
transition form factors $f$,$ g$, and $a_+$ are obtained by analytic
continuation from the spacelike region.  The form factor $a_-$ cannot
be obtained in this frame unless other components of the current
besides $\la J_{V-A}^+ \ra$ are calculated.  In the purely longitudinal
$q^+ > 0$ frame, all four form factors ($f$, $g$, and $a_\pm$) are
found from $\la J_{V-A}^+ \ra$ but the nonvalence contributions should
be computed in addition to the valence ones.  We confirmed that all
four form factors obtained in LFD are identical to the result of the
manifestly covariant calculation and the DYW results for $f$, $g$, and
$a_+$ are identical to those obtained in the purely longitudinal $q^+ >
0$ frame.

In our analysis, we do not find any zero-mode contribution to the
transition form factor $f(q^2)$ (or equivalently the axial-vector form
factor $A_1 (q^2)$). The absence of a zero-mode is not affected by the
modification of the vector meson vertex from $\Gamma^\mu = \gamma^\mu$ to
$\Gamma^\mu_{\rm LFQM}$.

For the numerical computation, we fixed the model parameters using the
normalization constraints in the elastic form factors and the available
experimental data of decay constants of the pseudoscalar ($B$) and
vector ($D^*,\rho$) mesons. Comparing the results of heavy-to-light
($B\to\rho$) and heavy-to-heavy ($B\to D^*$) transition form factors,
we find that the nonvalence contributions are significantly reduced in
the heavy-to-heavy results. Our results for the $B\to D^* l \nu_l$
decay process satisfy the constraints imposed by the flavor
independence on the heavy-to-heavy semileptonic decays~\cite{AW}.

\appendix

\section{Kinematics}
\label{AppA}

In this appendix we discuss in some detail the different reference
systems we used. In our previous publication\cite{BCJ}, we used the 
target-rest
frame (TRF), the Breit frame (BRT), and the Drell-Yan-West frame (DYW).
In the present case, where the momentum transfer is timelike, the TRF is
still straightforward to define, but the other frames are not. That is why we
give the detailed formulas here. We write the momenta in the LFD form:
$P = (P^+, P_x, P_y, P^-)$ with $P^2 = P^+ P^- - \vec{P}^{\,2}_\bot$.

\subsection{Target-rest frame}
\label{AppA1}
The momentum of the inital pseudoscalar meson with mass $M_1$ is
\begin{equation}
 P_1 = \left(M_1, 0,0, M_1 \right).
{\label{eq.A01}}
\end{equation}
If $M_2$ is the mass of the vector meson, $m^2_l$ is the invariant mass
square of the lepton pair in the final state, and $q$ is the
four-momentum transfer, the kinematical range of $q^2$ is
\begin{equation}
 m^2_l \leq q^2 \leq (M_1 - M_2)^2.
{\label{eq.A02}}
\end{equation}
Four-momentum conservation allows us to determine the kinematical range of
the three-momentum transfer. 

We write for $q$
\begin{equation}
 q = (q^+, \vec{q}_\bot, q^-)
{\label{eq.A03}}
\end{equation}
and write
\begin{equation}
 \vec{q}_\bot = Q\sin\theta \hat{n} = Q(\sin\theta\,\cos\phi, \sin\theta\, 
\sin\phi).
{\label{eq.A.04}}
\end{equation}
We define the quantity $M^2_q$ as follows
\begin{equation}
 M^2_q = M^2_1 - M^2_2 + q^2
{\label{eq.A05}}
\end{equation}
and find for the square of the length of the three-momentum transfer
\begin{equation}
 Q^2 = \frac{M^4_q - 4 M^2_1 q^2}{4 M^2_1}.
{\label{eq.A06}}
\end{equation}
The complete expression for $q$ is
\begin{equation}
 q = \left( \frac{M^2_q + 2M_1 Q \cos\theta}{2M_1}, Q\sin\theta\hat{n},
            \frac{M^2_q - 2M_1 Q \cos\theta}{2M_1} \right).
{\label{eq.A07}}
\end{equation}

\begin{figure}[t]
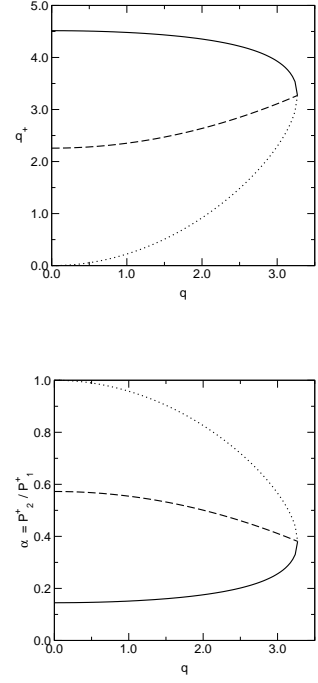

\psfig{figure=TRF_qplus_q.eps,height=4cm,width=4cm}

\vspace{7ex}

\psfig{figure=TRF_alpha_q.eps,height=4cm,width=4cm}
\caption{The quantities $q^+$ (top) and $\alpha = P^+_2/P^+_1$ (bottom)
in the TRF for $\theta = 0$ (solid), $\pi/2$ (dashed), and $\pi$  (dotted),
respectively, plotted for $q = \surd q^2$ from 0 to $(M_1 - M_2)^2$.   ($B \to D^*$)
{\label{fig.TRF}}}
\end{figure}

The behaviour of both $q^+$ and $\alpha$ is smooth as can be seen in 
Fig.~\ref{fig.TRF}.

\begin{figure}[t]

\vspace{3ex}

\psfig{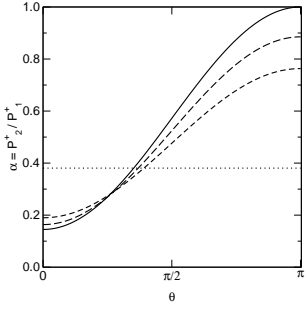}
\caption{The quantity $\alpha = P^+_2/P^+_1$ 
in the TRF for $q = 0$ (solid), $(M_1 - M_2)/2$ (long dashed), 
$(M_1 - M_2)/\surd 2$ (short dashed) and $M_1 - M_2$  (dotted),
respectively, plotted for $\theta$ from 0 to $\pi$.   ($B \to D^*$)
{\label{fig.TRFt}}}
\end{figure}

\subsection{Breit frame}
\label{AppA2}
The Breit frame is usually defined by the requirement that there is no
energy transfer. In the case of the elastic form factors this could be
achieved easily. However, for a time-like momentum $q$ the component
$q^0$ is not allowed to vanish in the physical region. One may define a
Breit-like frame in either of the two following ways.\\

\noindent
$(i)$ Real momenta\\
\begin{equation}
 P_1 = P + q/2, \quad P_2 = P - q/2.
{\label{eq.A08}}
\end{equation}
For $q^0 = 0$ and $\vec{P} = 0$ this choice of momenta corresponds to a
particle with momentum $\vec{q}/2$ bouncing off a `brick wall' and changing
its momentum to $-\vec{q}/2$. This process is only possible if the particle
with momentum $P_1$ has the same mass as the one with momentum $P_2$.

Our generalization drops the condition $q^0 = 0$. Then different
masses, $M_1 \neq M_2$, are allowed. Keeping $\vec{P} = 0$ simplifies
the formulas. One may relax the latter condition by a simple boost to a
frame where $\vec{P} \neq 0$.

The values of $P^0$ and $Q = |\vec{q}|$ that correspond to the on-shell
conditions $P^2_1 = M^2_1$ and $P^2_2 = M^2_2$ are given by
\bea
 P^0 &=& \sqrt{\frac{M^2_1 + M^2_2}{2} - \frac{q^2}{4}}, 
\nonumber\\
 Q &=& \sqrt{\frac{q^4 - 2(M^2_1 + M^2_2)\, q^2 + (M^2_1 - M^2_2)^2}
                {2(M^2_1 + M^2_2) - q^2}}.
{\label{eq.A09}}
\eea

The LF momenta are easily obtained. As we rely on $q^2 > 0$ and real
momenta, it is clear that $q^+ > 0$. We have

\begin{eqnarray}
 q^+ & = & \sqrt{q^2 + Q^2} + Q \cos\theta, \nonumber \\
 \vec{q}_\bot & = & Q \sin\theta\, \hat{n},\nonumber \\
 q^- & = & \frac{Q^2 \sin^2\theta + q^2}{\sqrt{q^2 + Q^2} + Q \cos\theta} .
{\label{eq.A10}}
\end{eqnarray}
Clearly, $q^+$ cannot vanish for real $Q$.\\

\begin{figure}
\psfig{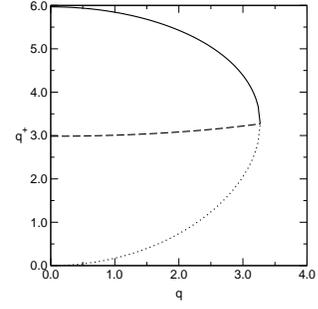} 

\vspace{7ex}

\psfig{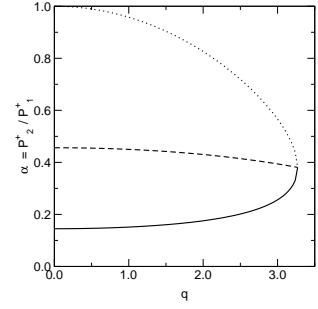}
\caption{The quantities $q^+$ (upper) and $\alpha = P^+_2/P^+_1$ (lower)
in the Breit frame with real momenta for $\theta = 0$ (solid), 
$\pi/2$ (dashed), and $\pi$  (dotted),
respectively, plotted for $q^2$ from 0 to $(M_1 - M_2)^2$.   ($B \to D^*$)
{\label{fig.BRT}}}
\end{figure}
The behaviour of both $q^+$ and $\alpha$ is smooth as can be seen 
in Fig.~\ref{fig.BRT}.

\begin{figure}

\vspace{3ex}

\psfig{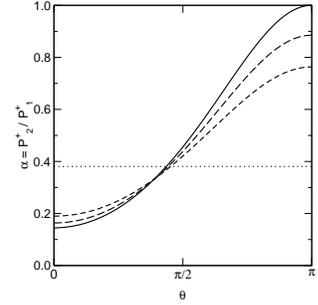}
\caption{The quantity $\alpha = P^+_2/P^+_1$ 
in the Breit frame for $q = 0$ (solid), $(M_1 - M_2)/2$ (long dashes), 
$(M_1 - M_2)/\surd 2$ (dotted) and $M_1 - M_2$  (dashed),
respectively, plotted for $\theta$ from 0 to $\pi$.   ($B \to D^*$)
{\label{fig.BRTt}}}
\end{figure}

\noindent
$(ii)$ Complex $q$\\
In order to avoid confusion we reserve the notation with
$Q$ for the case of real momenta.
In order to follow Ref.~\cite{BCJ} as close as possible we define
\begin{equation}
 q = \left( q \cos\theta, i\, q\sin\theta\,\hat{n},
            q \cos\theta \right).
{\label{eq.A11}}
\end{equation}
Next we determine $P$. Now we take $\vec{P} = 0$, but we allow for
$P^0 \neq 0$, otherwise we shall not be able to satisfy the on-shell
conditions for $P_1$ and $P_2$. Then, $P^2_1 = M^2_1$ and $P^2_2 = M^2_2$
give the equations
\begin{eqnarray}
 P^+ + P^- & = & \frac{M^2_1 - M^2_2}{q\cos\theta},
 \nonumber \\
 P^+ P^- & = & \frac{M^2_1 + M^2_2}{2} - \frac{q^2}{4}.
{\label{eq.A12}}
\end{eqnarray}
In the kinematically allowed domain both $P^+ + P^-$ and $P^+ P^-$ are
positive, so both are separately positive. We find for them
\begin{eqnarray}
 P^+ & = & \frac{M^2_1 - M^2_2}{2q\cos\theta}
\nonumber\\
&& + \frac{1}{2} \sqrt{\frac{(M^2_1 - M^2_2)^2}{q^2 \cos^2\theta} -
                             2(M^2_1 + M^2_2) + q^2}, 
\nonumber \\
 P^- & = & \frac{M^2_1 - M^2_2}{2q\cos\theta}
\nonumber\\
&& - \frac{1}{2} \sqrt{\frac{(M^2_1 - M^2_2)^2}{q^2 \cos^2\theta} -
                             2 (M^2_1 + M^2_2) + q^2}.
\nonumber\\
{\label{eq.A13}}
\end{eqnarray}
For this unphysical kinematics $q^+ = 0$ is allowed. The lower bound
$q^2 = 0$ leads to a divergent limit for $P^+$, while $P^-$ tends to 0. 
Their product is of course finite for all values of $q$.

\begin{figure}[t]
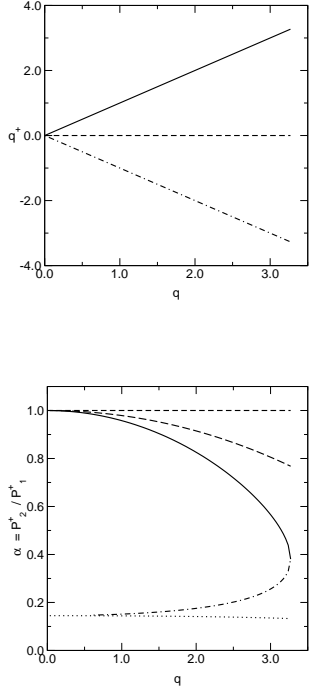

\psfig{figure=BRTIqplus_q.eps,height=4cm,width=4cm} 

\vspace{8ex}

\psfig{figure=BRTIalpha_q.eps,height=4cm,width=4cm}
\caption{The quantities $q^+$ (upper) and $\alpha = P^+_2/P^+_1$
(lower) in the Breit frame with complex momenta for $\theta = 0$
(solid), $\pi/4$ (long dashed), $\pi/2$ (short dashed), $3\pi/4$
(dotted), and $\pi$  (dot dashed), respectively, plotted for $q^2$ from
0 to $(M_1 - M_2)^2$. ($B \to D^*$)
{\label{fig.BRTI}}}
\end{figure}

\begin{figure}

\vspace{3ex}

\psfig{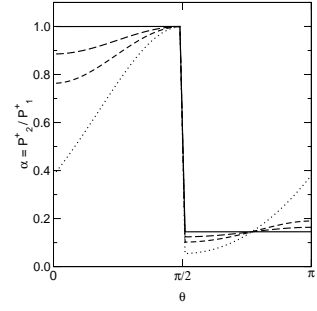}
\caption{The quantity $\alpha = P^+_2/P^+_1$ in the Breit frame with
complex momenta for $q = 0$ (solid), $(M_1 - M_2)/2$ (long dashes),
$(M_1 - M_2)/\surd 2$ (short dashed) and $M_1 - M_2$  (dotted),
respectively, plotted for $\theta$ from 0 to $\pi$.   ($B \to D^*$)
\label{fig.BRTIt}}
\end{figure}

The behaviour of $q^+$ is smooth (linear), but $\alpha$ has a
singularity at $\theta = \pi/2$. This singularity is a branch point.
For values of $\theta$ between 0 and $\pi/2$, $\alpha$ increases to 1
for all values of $q$.  In the interval $[\pi/2, \pi]$ $\alpha$
increases for all $q$ from a value of 
$(4 M^2_2 - q^2)/(4 M^2_1 - q^2)$ to its value at
$\theta = \pi$. This behaviour is illustrated in Fig.~\ref{fig.BRTIt}.

\subsection{Drell-Yan-West frame}
\label{AppA3}
As the DYW-frame is characterized by $q^+ = 0$, we are obliged to take
$\vec{q}_\bot$ purely imaginary to get $q^2 > 0$. The solution of the
on-shell conditions is particularly simple. Our final results are
\begin{eqnarray}
 q & = & \left(0, i q \hat{n}, \frac{M^2_1 - M^2_2 + q^2}{P^+_1} \right),
 \nonumber \\
 P_1 & = & \left( P^+_1, \vec{0}_\bot, \frac{M^2_1}{P^+_1} \right),
 \nonumber \\
 P_2 & = & \left( P^+_1, -i q \hat{n}, \frac{M^2_2- q^2}{P^+_1}  \right).
{\label{eq.A14}}
\end{eqnarray}

If we substitute for the arbitrary $P^+_1$ the value $M_1$, we
obtain a quasi TRF kinematics. Needless to say that this kinematics cannot
be obtained from the formulas given before for the TRF.

\section{Elastic form factors and decay constants of mesons with unequal 
quark masses} 
\label{AppB}

In this appendix, we summarize the manifestly covariant formulae of 
elastic form factors and decay constants of pseudoscalar and vector mesons
with the unequal quark masses such as $B$ and $D^*$ mesons.

\subsection{Pseudoscalar Meson Electromagnetic Form Factor}
\label{AppB1}
The electromagnetic form factor $F_{\rm ps}(q^2)$ of pseudoscalar meson is
defined by the matrix element given by \begin{equation} \la P'|J^\mu|P
\ra = (P'^\mu + P^\mu ) F_{\rm ps}(q^2), \end{equation} where $P$ and $P'=
P +q$ are the four-momenta of initial and final states, respectively.
If the meson is made of a quark and an antiquark with mass(charge) values 
$m_1(e_1)$ and $m_2(e_2)$, respectively, $F_{\rm ps}(q^2)$ is given by
\begin{widetext}
\bea
\label{psff}
F_{\rm ps}(q^2) &=&
\frac{g_{\rm ps}^2 \Lambda_1^4 e_1}{8\pi^2(\Lambda_1^2 - m_1^2)^2 }
\int_0^1 dx \int_0^{1-x} dy
\biggl[\{4-3(x+y)\}\ln\frac{C_{m_2 \Lambda_1 m_1} C_{m_2 m_1 \Lambda_1}}
{C_{m_2 \Lambda_1 \Lambda_1} C_{m_2 m_1 m_1}} \nonumber \\
&&+\{-(x+y)(x+y-1)^2 M^2 -(2-x-y)xyq^2 + 2(x+y-1)m_1 m_2
-(x+y)m_1^2\}C_{12}\biggr] \nonumber \\
&&+ (1 \leftrightarrow 2 ),
\eea
\end{widetext}
where $e_1 + e_2$ must be indentical to the charge of the meson,
\bea
C_{12} = \frac{1}{C_{m_2 \Lambda_1 \Lambda_1}}-
\frac{1}{C_{m_2 \Lambda_1 m_1}}-\frac{1}{C_{m_2 m_1 \Lambda_1}}
+\frac{1}{C_{m_2 m_1 m_1}}
\nonumber\\
\label{eq.B3}
\eea
and 
\begin{eqnarray}
\label{coeff}
C_{m_2 \Lambda_1 \Lambda_1}&=&(x+y)(1-x-y)M^2 + xyq^2
\nonumber\\
&&-(x+y)\Lambda_1^2 -(1-x-y)m_2^2, \nonumber \\
C_{m_2 \Lambda_1 m_1}&=&(x+y)(1-x-y)M^2 + xyq^2
\nonumber\\
&&-(x\Lambda_1^2 + y{m_1}^2) -(1-x-y)m_2^2, \nonumber \\
C_{m_2 m_1 \Lambda_1}&=&(x+y)(1-x-y)M^2 + xyq^2
\nonumber\\
&&-(x{m_1}^2+y\Lambda_1^2) -(1-x-y)m_2^2, \nonumber \\
C_{m_2 m_1 m_1}&=&(x+y)(1-x-y)M^2 + xyq^2
\nonumber\\
&& -(x+y){m_1}^2 -(1-x-y)m_2^2.
\end{eqnarray}

\subsection{Pseudoscalar Meson Decay Constant}
\label{AppB2}
The decay constant $f_{\rm ps}$ of a pseudoscalar meson is defined by
the matrix element,
\begin{equation}\label{eq:B5}
\la 0 |J^\mu_{V-A}|P \ra = i P^\mu f_{\rm ps}.
\end{equation}
From this definition, we find
\begin{eqnarray}\label{eq:B6}
f_{\rm ps} &=& \frac{g_{\rm ps} \Lambda_1^2
\Lambda_2^2}{4\pi^2(\Lambda_1^2-m_1^2)(\Lambda_2^2-m_2^2)}
\int_0^1 dx 
\nonumber\\
&&\times\{ x m_1 + (1-x) m_2 \}
\ln\frac{C_{m_1 \Lambda_2} C_{\Lambda_1 m_2}}
{C_{\Lambda_1 \Lambda_2} C_{m_1 m_2}},
\end{eqnarray}
where
\begin{eqnarray}
\label{coef}
C_{m_1 m_2} &=& x(1-x)M^2 -x m_1^2 -(1-x)m_2^2, \nonumber \\
C_{m_1 \Lambda_2} &=& x(1-x)M^2 -x m_1^2 -(1-x)\Lambda_2^2, \nonumber \\
C_{\Lambda_1 m_2} &=& x(1-x)M^2 -x \Lambda_1^2 -(1-x)m_2^2, \nonumber \\
C_{\Lambda_1 \Lambda_2} &=& x(1-x)M^2 -x \Lambda_1^2 
-(1-x)\Lambda_2^2.
\end{eqnarray}

\subsection{Vector Meson Electromagnetic Form Factors}
\label{AppB3}
The electromagnetic form factors ($F_1(q^2)$, $F_2(q^2)$, and $F_3(q^2)$)
of a vector meson are defined by the matrix element between the 
initial state of helicity $h$ and four-momentum $P$ and the final 
state of $h'$ and $P'$:
\begin{eqnarray}
\la P',h'|J^\mu|P,h \ra &=&
-\epsilon^*_{h'} \epsilon_h (P'+ P)^\mu F_1(q^2)
\nonumber\\
&&+(\epsilon^\mu_h q\cdot \epsilon^*_{h'}-\epsilon^{*\mu}_{h'} q\cdot
\epsilon_{h} F_2(q^2) \nonumber \\
&&+ \frac{(\epsilon^*_{h'}\cdot q)(\epsilon_h \cdot q)}{2M^2}(P' + P)^\mu
F_3(q^2),
\nonumber\\
\end{eqnarray}
where $\epsilon_h$($\epsilon^*_{h'}$) is the polarization vector of the
initial(final) helicity $h$($h'$) state.  If the meson is made of a
quark and an antiquark with mass(charge) values $m_1(e_1)$ and
$m_2(e_2)$, respectively, $F_i(q^2)$($i=1,2,3$) are given by
\begin{widetext}
\begin{eqnarray}
\label{ff1}
F_1(q^2) &=& \frac{g_{\rm v}^2 \Lambda_1^4 e_1}{8\pi^2(\Lambda_1^2 -
m_1^2)^2} \int_0^1 dx \int_0^{1-x} dy
\biggl[(2-x-y)\ln\frac{C_{m_2 \Lambda_1 m_1} C_{m_2 m_1 \Lambda_1}}
{C_{m_2 \Lambda_1 \Lambda_1} C_{m_2 m_1 m_1}} \nonumber \\
&&+\{-(x+y)(x+y-1)^2 M^2 -(2-x-y)xyq^2
+ 2(x+y-1)m_1 m_2 -(x+y)m_1^2\}C_{12}\biggr] \nonumber \\
&&+ (1 \leftrightarrow 2 ),
\\
F_2(q^2) &=& -\frac{g_{\rm v}^2 \Lambda_1^4 e_1}{8\pi^2(\Lambda_1^2 -
m_1^2)^2} \int_0^1 dx \int_0^{1-x} dy
\biggl[(2+x+y)\ln\frac{C_{m_2 \Lambda_1 m_1} C_{m_2 m_1 \Lambda_1}}
{C_{m_2 \Lambda_1 \Lambda_1} C_{m_2 m_1 m_1}} \nonumber \\
&&+\{(x+y)(x+y+1)(x+y-1) M^2 -xy(x+y)q^2
-(x+y){m_1}^2- 2m_1m_2\}C_{12}\biggr] \nonumber \\
&+& (1 \leftrightarrow 2 ),
\\
F_3(q^2) &=& \frac{g_{\rm v}^2  \Lambda_1^4 e_1}{8\pi^2(\Lambda_1^2 -
m_1^2)^2} \int_0^1 dx \int_0^{1-x} dy \;
8xy(x+y-1)M^2 C_{12}
+ (1 \leftrightarrow 2 ),
\end{eqnarray}
\end{widetext}
where $e_1 + e_2$ must be equal to the charge of the meson and
$C_{12}$ is identical to the one given in Eq.~(\ref{eq.B3}).

\subsection{Vector Meson Decay Constant}
\label{AppB4}
The decay constant $f_{\rm v}$ of a vector meson is defined by
the matrix element,
\begin{equation}\label{eq:B12}
\la 0 |J^\mu_{V-A}|P,h \ra = i M f_{\rm v} \epsilon^\mu(h),
\end{equation}
where $\epsilon(h)$ is the polarization vector of helicity $h$ state.
From this definition, we find
\begin{eqnarray}\label{eq:B13}
f_{\rm v} &=& \frac{g_{\rm v} \Lambda_1^2
\Lambda_2^2}{4\pi^2 M(\Lambda_1^2-m_1^2)(\Lambda_2^2-m_2^2)}
\int_0^1 dx \nonumber \\
&&\times\biggl[ \{m_1 m_2 + x(1-x) M^2 \}
\ln\frac{C_{m_1 \Lambda_2} C_{\Lambda_1 m_2}}
{C_{\Lambda_1 \Lambda_2} C_{m_1 m_2}} \nonumber \\
&-& C_{\Lambda_1 \Lambda_2} \ln (-C_{\Lambda_1 \Lambda_2})
+ C_{m_1 \Lambda_2} \ln (-C_{m_1 \Lambda_2}) \nonumber \\
&+& C_{\Lambda_1 m_2} \ln (-C_{\Lambda_1 m_2})
- C_{m_1 m_2} \ln (-C_{m_1 m_2})\biggr],
\nonumber\\
\end{eqnarray}
where $C_{m_1 m_2}$, $C_{m_1 \Lambda_2}$, $C_{\Lambda_1 m_2}$,
$C_{\Lambda_1\Lambda_2}$ are given by Eq.~(\ref{coef}).

\section{Analytic expressions of $g(q^2)$, $a_+(q^2)$, and
$f(q^2)$ in the $q^+=0$ frame for $\Gamma^+=\gamma^+$}
\label{AppC}

For the numerical analysis of the weak form factors in the $q^+=0$
frame, we use Feynman parameterization to integrate out the transverse
momentum, ${\bf k}_\perp$.

Similar to the covariant analysis, we first separate the 
energy denominators as follows:
\bea{\label{anal:1}}
&&\frac{1}{(M^2_1 - M^2_0)(M^2_1-M^2_{\Lambda_1})
(M^2_2-M'^2_0)(M^2_2-M'^2_{\Lambda_2})}
\nonumber\\
&&=\frac{(1-x)^2}{(m^2_1-\Lambda^2_1)(m^2_2-\Lambda^2_2)}
\biggl(\frac{1}{N_1}-\frac{1}{N_{1\Lambda}}\biggr)
\biggl(\frac{1}{N_2}-\frac{1}{N_{2\Lambda}}\biggr)
\nonumber\\
\eea
where $N_1=M^2_1-M^2_0$, $N_{1\Lambda}=M^2_1-M^2_{\Lambda_1}$,
$N_2=M^2_2-M'^2_0$, and 
$N_{2\Lambda}=M^2_2-M'^2_{\Lambda_2}$.

Now using Feynman parametrization 
\bea{\label{ab}}
\frac{1}{ab}&=&
\int^1_0 \frac{dx}{[a x + b(1-x)]^2}
\eea
we obtain from Eqs.~(\ref{g_val}), (\ref{ap_val}), and (\ref{f_val})
\begin{widetext}
\bea{\label{g_anal}}
g(q^2)&=&\frac{{\cal N}}{8\pi^2}
\int^1_0 x dx\int^1_0 dy
\biggl[{\cal A}_p-x(1-y)(m_1-m_2)\biggr]
\nonumber\\
&&\times
\biggl\{
\frac{1}{a_1 y + a_2(1-y) + y(1-y)x^2 q^2}
- (a_2\to a_{2\Lambda}) - (a_1\to a_{1\Lambda})
+ (a_1\to a_{1\Lambda},a_2\to a_{2\Lambda})
\biggr\},\\
a_+(q^2)&=&\frac{{\cal N}}{8\pi^2}
\int^1_0 x dx\int^1_0 dy
\biggl[(1-2x){\cal A}_p-x(1-y)[(1-2x)m_1-m_2-2(1-x)m]\biggr]
\nonumber\\
&&\times
\biggl\{
\frac{1}{a_1 y + a_2(1-y) + y(1-y)x^2 q^2}
- (a_2\to a_{2\Lambda}) - (a_1\to a_{1\Lambda})
+ (a_1\to a_{1\Lambda},a_2\to a_{2\Lambda})
\biggr\},
\\
f(q^2)&=& (M^2_1-M^2_2 - q^2)a_+(q^2)
- \frac{{\cal N}}{4\pi^2}
\int^1_0 dx\int^1_0 dy\biggl\{
C_1\biggl[
\frac{1}{a_1 y + a_2(1-y) + y(1-y)x^2 q^2} - (a_2\to a_{2\Lambda}) 
\nonumber\\
&&\hspace{2cm}- (a_1\to a_{1\Lambda})
+ (a_1\to a_{1\Lambda},a_2\to a_{2\Lambda})
\biggr]
- C_2\ln\biggl(\frac{a_{12\Lambda}a_{1\Lambda 2}}
{a_{12}a_{1\Lambda 2\Lambda}}\biggr)
\biggr\}
\eea
\end{widetext}
where
\bea{\label{a12}}
a_1&=& x(1-x)M^2_1-xm^2_1-(1-x)m^2
\nonumber\\
a_2&=& x(1-x)M^2_2 - xm^2_2-(1-x)m^2
\nonumber\\
a_{1\Lambda}&=&a_1(m_1\to\Lambda_1),
a_{2\Lambda}=a_2(m_2\to\Lambda_2),
\eea
\bea{\label{C12}}
C_1&=&{\cal A}_p[x(1-x)M^2_2 + m_2m - x^2 q^2]
\nonumber\\
&&- x^2(1-y)^2(xm_1+m_2-xm) q^2
\nonumber\\
&&+ x^2(1-y)[2x(m_1-m)+m_2+m]q^2
\nonumber\\
C_2&=& x(m_1-m) + m_2, 
\eea
and 
\bea{\label{a12L}}
a_{12}&=&a_1 y + a_2 (1-y)+y(1-y)x^2 q^2
\nonumber\\
a_{12\Lambda}&=&a_1 y+a_{2\Lambda}(1-y) + y(1-y) x^2 q^2
\nonumber\\
a_{1\Lambda2}&=&a_{1\Lambda} y + a_2 (1-y)+y(1-y)x^2 q^2
\nonumber\\
a_{1\Lambda2\Lambda}&=&a_{1\Lambda} y 
+ a_{2\Lambda} (1-y)+y(1-y)x^2 q^2
\eea

\section{Form factors $g(q^2)$, $a_+(q^2)$, and $f(q^2)$ in the $q^+=0$
frame for $\Gamma^+_{\rm LFQM}$} 
\label{AppD}

In this appendix, we give the exact LF expressions for the form factors
$g(q^2),a_+(q^2)$, and $f(q^2)$ in the $q^+=0$ frame for the more
realistic LF vector meson vertex function given by Eq.~(\ref{RV:1}).
We shall write $g^{\rm LFQM}(q^2)$, $a^{\rm LFQM}_+(q^2)$, and $f^{\rm
LFQM}(q^2)$ for the vertex $\Gamma^+_{\rm LFQM}$ to distinguish them
from those obtained for $\Gamma^+=\gamma^+$.

To obtain the form factor $g(q^2)$, we first calculate the trace
$T^{+(h=1)}_V$ for the vector current with transverse polarization,
which is given by
\bea\label{eq:D1}
T^{+(h=1)}_V &=&- 4i\frac{(p_2-k)\cdot\epsilon^*(h=1)}{M'_0+m_2+m}
\nonumber\\
&&\;\;\times\epsilon^{+\mu\nu\sigma}(p_{1\rm on})_\mu(p_{2\rm on})_\nu
(k_{\rm on})_\sigma
\nonumber\\
&=&-4\sqrt{2}P^+_1\epsilon^{+-xy}
\frac{{\bf k}^2_\perp q^L - ({\bf k}_\perp\cdot{\bf q}_\perp)k^L}
{M'_0 + m_2 + m},
\nonumber\\
\eea 
where we use the identity $({\bf k}_\perp\times{\bf q}_\perp)_z
=i(k^Rq^L-{\bf k}_\perp\cdot{\bf q}_\perp)$.
Note that $T^{+(h=1)}_V$ is independent of $k^-$ (which is due to
$\epsilon^+(h=1)=0$) and thus free from
the zero-mode contribution as in the case of $\Gamma^+=\gamma^+$.

Modifying $S^{+(h=1)}_{{\rm on}\;V}\to S^{+(h=1)}_{{\rm 
on}\;V}-T^{+(h=1)}_V$ in Eq.~(\ref{ch:4}) for 
$\Gamma^+$ given by Eq.~(\ref{RV:1}), we obtain
the form factor $g^{\rm LFQM}(q^2)$ in the $q^+=0$ frame as
\begin{widetext}
\bea{\label{eq:D2}}
g^{\rm LFQM}(q^2)&=& 
-\frac{g_1g_2\Lambda^2_1\Lambda^2_2}{(2\pi)^3}\int^1_0
\frac{dx}{x(1-x)^4}\int d^2{\bf k}_\perp
\frac{1}{(M^2_1 - M^2_0)(M^2_1-M^2_{\Lambda_1})
(M^2_2-M'^2_0)(M^2_2-M'^2_{\Lambda_2})}
\nonumber\\
&&\times\biggl\{ {\cal A}_P +
\frac{{\bf k}_\perp\cdot{\bf q}_\perp}{{\bf q}^2_\perp}(m_1 - m_2)
+ \frac{2}{M'_0+m_2+m}
\biggl[{\bf k}^2_\perp -
\frac{({\bf k}_\perp\cdot{\bf q}_\perp)^2}{{\bf q}^2_\perp}
\biggr] \biggr\}.
\eea
\end{widetext}
We note that our result for $g^{\rm LFQM}(q^2)$ in Eq.~(\ref{eq:D2}) is 
equivalent to that obtained by Jaus~\cite{Jaus} (see, for example,
Eq.~(4.13) in Ref.~\cite{Jaus}) and free from the zero-mode contribution.

Now, the trace $T^{+(h)}_A$ for the axial-vector current is given by
\bea\label{eq:D3}
T^{+(h)}_A &=& 4\frac{(p_2-k)\cdot\epsilon^*(h)}{M'_0+m_2 + m}
\biggl[ (p_{2\rm on}\cdot k_{\rm on}-m_2m)p^+_1
\nonumber\\
&+& (p_{1\rm on}\cdot k_{\rm on} + m_1 m)p^+_2
 - (p_{1\rm on}\cdot p_{2\rm on} + m_1 m_2)k^+
\nonumber\\
&+& (k^- -k^-_{\rm on})p^+_1p^+_2 \biggr].
\eea
The form factor $a_+^{\rm LFQM}(q^2)$ in the $q^+=0$ frame is obtained
from the axial-vector current with transverse polarization ($h=1$)
(see the $\alpha\to 1$ limit in Eq.~(\ref{ch:5})).
Explicitly, the trace is given by
\bea\label{eq:D4}
T^{+(h=1)}_A&=& -\frac{4\sqrt{2}P^+_1}{x(M'_0+m_2 +m)}
(x q^L + k^L)
\nonumber\\
&\times& \biggl\{
{\bf k}_\perp\cdot{\bf k'}_\perp
+ [(1-x)m-xm_2]{\cal A}_p
\nonumber\\
&&\;\; +\;
x(1-x)^2(k^- - k^-_{\rm on})P^+_1 \biggr\}.
\eea
Even though $(p_2-k)\cdot\epsilon^*(h=1)$ is independent of $k^-$,
there is a possibility to get a zero-mode contribution from the last
term, {\it i.e.}, the term proportional to $(k^--k^-_{\rm on})$, in
Eq.~(\ref{eq:D3}) or ~(\ref{eq:D4}). While the valence part,
$[T^{+(h=1)}_A]_{\rm val}$, is obtained for $k^-=k^-_{\rm on}$, the
zero-mode part, $[T^{+(h=1)} _A]_{\rm zm}$, is obtained for
$k^-=k^-_{m_1}$ or $k^-=k^-_{\Lambda_1}$.  
However, counting only the longitudinal momentum fraction terms, one can
easily find from Eq.~(\ref{eq:D4}) that the zero-mode part of
$T^{+(h=1)}_A$ vanishes as $[T^{+(h=1)}_A]_{zm}\sim (1-x)^{3/2}$ in the
$\alpha\to 1$ (or $x\to 1$) limit and the valence part as
$\sim\sqrt{1-x}$ in the $x\to 1$ limit.
Therefore, following the argument given in
Sec.~\ref{SecIIID}, there is no zero-mode contribution to the form factor
$a^{\rm LFQM}_+(q^2)$.  Modifying $S^{+(h=1)}_{{\rm on}\;A}\to
S^{+(h=1)}_{{\rm on}\;A} -[T^{+(h=1)}_A]_{\rm val}$ in Eq.~(\ref{ch:5})
and taking the limit of $\alpha\to 1$, we obtain
\begin{widetext}
\bea{\label{eq:D5}}
a^{\rm LFQM}_+(q^2)&=& 
-\frac{g_1g_2\Lambda^2_1\Lambda^2_2}{(2\pi)^3}\int^1_0
\frac{dx}{x(1-x)^4}\int d^2{\bf k}_\perp
\frac{1}{(M^2_1 - M^2_0)(M^2_1-M^2_{\Lambda_1})
(M^2_2-M'^2_0)(M^2_2-M'^2_{\Lambda_2})}
\nonumber\\
&&\times\biggl\{ (1-2x){\cal A}_P +
\frac{{\bf k}_\perp\cdot{\bf q}_\perp}{{\bf q}^2_\perp}
\;[\; (1-2x) m_1 - m_2 - 2(1-x) m \;]
\nonumber\\ 
&&\hspace{0.5cm}
-\frac{2(x{\bf q}^2_\perp + {\bf k}_\perp\cdot{\bf q}_\perp)}
{x{\bf q}^2_\perp(M'_0+m_2 +m)}
[{\bf k}_\perp\cdot{\bf k'}_\perp
+ ( (1-x)m -x m_2){\cal A}_P]
\biggr\},
\eea
\end{widetext}
which is again equivalent to that obtained by Jaus~\cite{Jaus} (see,
for example, his Eq.~(4.14)).

Finally, we need to compute the trace in Eq.~(\ref{eq:D3}) with the
longitudinal polarization vector ($h=0$) to obtain $f^{\rm LFQM}(q^2)$.
Here we again separate the trace term, $T^{+(h=0)}_A$, into the
valence part, $[T^{+(h=0)}_A]_{\rm val}$, with $k^-=k^-_{\rm on}$ and
the possible zero-mode part, $[T^{+(h=0)}_A]_{\rm zm}$, with 
$k^-=k^-_{m_1}$
and $k^-=k^-_{\Lambda_1}$. Explicitly, the valence part is given by
\bea\label{eq:D6}
[T^{+(h=0)}_A]_{\rm val}&=&-\frac{4P^+_1}{xM_2(M'_0+m_2+m)}
 \nonumber \\
 & \times & \{xM^2_2 + (1-x)M'^{2}_0\}
\nonumber\\
&\times&\{{\bf k}_\perp\cdot{\bf k'}_\perp
+ [(1-x)m - xm_2]{\cal A}_P\},
\nonumber\\
\eea
while the possible zero-mode part for $k^-=k^-_{\Lambda_1}$ is given by
\bea\label{eq:D7}
[T^{+(h=0)}_A]_{\rm zm}&=&\frac{4P^+_1}{xM_2(M'_0+m_2+m)}
\nonumber\\
&\times&\{x {\tilde M}^2_{\Lambda_1} + xM^2_2 -M^2_1 -{\bf q}^2_\perp\}
\nonumber\\
&\times&\{{\bf k}_\perp\cdot{\bf k'}_\perp
+ [(1-x)m - xm_2]{\cal A}_P
\nonumber\\
&&+x(1-x)^2(M^2_1-M^2_{\Lambda_1})
\},
\eea
where ${\tilde M}^2_{\Lambda_1}$ is defined as
\bea\label{eq:D8}
{\tilde M}^2_{\Lambda_1}
=\frac{\Lambda^2_1 + [{\bf k}_\perp -(1-x){\bf q}^2_\perp]}{x(1-x)}.
\eea
The zero-mode part for $k^-=k^-_{m_1}$ can
be easily obtained by changing $\Lambda_1\to m_1$
in Eq.~(\ref{eq:D7}).

Counting the longitudinal momentum fraction terms in Eq.~(\ref{eq:D7}),
one can easily find the singular behavior given by Eq.(\ref{RV:4}), {\it 
i.e.} 
\bea\label{eq:D9}
[T^{+(h=0)}_A]_{\rm zm}\sim\sqrt{\frac{1}{1-x}}
\eea
as $x\to 1$. However, as we showed in Sec.~\ref{SecIIIE} (see
Eq.(\ref{RV:5})), there is no zero-mode contribution to $f^{\rm
LFQM}(q^2)$ even though the trace term itself shows singular behavior
as $x\to 1$.

Therefore, the form factor $f^{\rm LFQM}(q^2)$ in the $q^+=0$ frame can be
obtained from the valence contribution only (see Eq.~(\ref{ch:4})) and
it is given by
\begin{widetext}
\bea\label{eq:D10}
f^{\rm LFQM}(q^2)
=\frac{M_2}{P^+_1}\la J^+_A\ra^{h=0}_{\rm LFQM}
- (M^2_1-M^2_2 +{\bf q}^2_\perp)a^{\rm LFQM}_+(q^2),
\eea
where
\bea{\label{eq:D11}}
\la J^+_A\ra^{h=0}_{\rm LFQM}&=&
\frac{g_1g_2\Lambda^2_1\Lambda^2_2}{(2\pi)^3}\int^1_0
\frac{dx}{x(1-x)^4}\int d^2{\bf k}_\perp
\frac{S^{+(h=0)}_{\rm on\;A}-[T^{+(h=0)}_A]_{\rm val}}
{(M^2_1 - M^2_0)(M^2_1-M^2_{\Lambda_1})
(M^2_2-M'^2_0)(M^2_2-M'^2_{\Lambda_2})}.
\eea
\end{widetext}
We note the difference from the conclusion drawn by
Jaus~\cite{Jaus,Jaus02}, where the author claimed that the form factor
$f(q^2)$ receives a zero-mode contribution.

\acknowledgments
This work was supported in part by a grant from the U.S.Department of
Energy (DE-FG02-96ER 40947) and the National Science Foundation
(INT-9906384). This work was started when HMC and CRJ visited the Vrije
Universiteit and they want to thank the staff of the department of
physics at VU for their kind hospitality.  The North Carolina
Supercomputing Center and the National Energy Research Scientific
Computer Center are also acknowledged for the grant of Cray time.

\end{document}